\documentclass[10pt,a4paper,notitlepage]{article}
\usepackage[activeacute,english]{babel}
\usepackage[latin1]{inputenc}
\usepackage[active]{srcltx}
\usepackage[toc,page]{appendix}
\usepackage{fancyhdr}
\usepackage{amsmath}
\usepackage{amsfonts}
\usepackage{amssymb}
\usepackage[mathcal]{euscript}
\usepackage{graphicx}
\usepackage{caption}
\usepackage{subcaption}
\usepackage{xcolor}
\usepackage[normalem]{ulem}
\usepackage{authblk}
\makeatletter
\let\old@startsection=\@startsection
\let\oldl@section=\l@section
\renewcommand{\@startsection}[6]{\old@startsection{#1}{#2}{#3}{#4}{#5}{#6\mathversion{bold}}}
\renewcommand{\l@section}[2]{\oldl@section{\mathversion{bold}#1}{#2}}
\makeatother

\makeatletter \@addtoreset{equation}{section} \makeatother

\title{\textbf{Ginzburg-Landau Approach\\
to Holographic Superconductivity}}

\author{Aldo Dector\thanks{Email address: aldo.dector@gmail.com, dector@ecm.ub.es}}

\affil{\textit{ Departament d'Estructura i Constituents de la Materia}\\
\textit{Facultat de F\'isica, Universitat de Barcelona}\\
\textit{Av. Diagonal 647, 08028 Barcelona, Spain}}

\date{}

\begin{document}
\begin{titlepage}
    \maketitle

    \begin{abstract}
		
We construct a family of minimal phenomenological models for holographic superconductors in $d=4+1$ $AdS$ spacetime and study the effect of scalar and gauge field fluctuations. By making a Ginzburg-Landau interpretation of the dual field theory, we determine through holographic techniques a phenomenological Ginzburg-Landau Lagrangian and the temperature dependence of physical quantities in the superconducting phase. We obtain insight on the behaviour of the Ginzburg-Landau parameter and whether the systems behaves as a Type I or Type II superconductor. Finally, we apply a  constant external magnetic field in a perturbative approach following previous work by D'Hoker and Kraus, and obtain droplet solutions which signal the appearance of the Meissner effect. 
		
    \end{abstract}
  \end{titlepage}

\clearpage

\tableofcontents

\newpage


\section{Introduction}

The $AdS/CFT$ correspondence \cite{Aharony:1999ti} has proven to be one of the most important recent developments in theoretical physics and, because of its strong/weak-coupling duality character, has become a useful tool for studying previously inaccessible physical systems where the usual perturbative approaches fail to apply. In this line of thought, one of the most interesting and promising applications of the $AdS/CFT$ correspondence is in the area of condensed matter physics, where holographic techniques are expected to shed some light on the study of systems where strong-coupling forbids a quasi-particle description, such as in high-temperature superconductors. These models of superconductivity through holographic methods, customarily called \textit{holographic superconductors}, have successfully reproduced some of the main features found in real-world superconductors, and form a rapidly growing area of study (see, e.g. \cite{Hartnoll:2008kx, Hartnoll:2009sz, Horowitz:2010gk}).

\bigskip

The study of the effect of external magnetic fields on holographic superconductors was addressed since the appearance of the first papers on holographic superconductivity. Most of the previous research was focused on $2+1$ dual field theories, motivated by possible applications to high temperature superconductors. However, it is of obvious interest to investigate the effect of magnetic fields in holographic models describing $3+1$ dimensional systems. Another important reason is that the breaking of the superconducting phase by probing a system with an external magnetic field provides one of the main ways of classifying a superconductor. Roughly speaking, a superconductor is said to be \textit{type I} when the system goes from the superconducting to the normal phase in a first order transition as the value of the external magnetic field is increased beyond a critical value $B_{c}$, all of this resulting in a separation between macroscopic regions of normal phase and superconducting phase in the material. On the other hand, a \textit{type II} superconductor has two critical values: below a first critical value $B_{c_{1}}$ the system is in a superconducting phase, but as the value of the field is increased, a stable vortex lattice (\textit{Abrikosov vortices}) begins to form inside the material where the magnetic field can penetrate until a second critical value $B_{c_{2}}$ is reached and the system enters fully in the normal phase. In this case the phase transitions are second order in $B$.

\bigskip

Starting from a minimal family of holographic superconducting models in the bulk, the main focus of this paper will be to implement a phenomenological description of the dual field theory in terms of Ginzburg-Landau theory using holographic techniques. We will show that the Ginzburg-Landau description captures the basic features of holographic superconductors. We start by adding small scalar field and gauge field perturbations to these bulk models. We find that we can consistently determine a phenomenological Ginzburg-Landau Lagrangian for the boundary theory, as well as the characteristic lengths of the system, and from this we can calculate the Ginzburg-Landau parameter $\kappa$ of the holographic superconductor. The value of this parameter will in turn allow us to determine whether the system is type I or II. Finally, we turn off the scalar and gauge perturbations and instead apply a constant magnetic field to our holographic superconductor by adding a finite magnetic component to the gauge field and constructing the gravity solutions up to second order in the magnetic field $B$. This is done using the black brane solution described in \cite{D'Hoker:2009bc}, which we exploit for the first time in the context of holographic superconductivity. Once we do this, we proceed to obtain droplet solutions of the system. 
  
	\bigskip
	
There is some evidence that holographic superconductors describing $2+1$ dimensional field theories mostly exhibit type II behaviour \cite{Hartnoll:2008kx, Domenech:2010nf}. The standard argument \cite{Hartnoll:2008kx} is that,  when applying an external $3+1$ dimensional magnetic field to a $2+1$ dimensional system, the free energy needed to expel it scales as the volume, while the free energy that the system gains from being in a superconducting state scales as the area. In this $2+1$ dimensional case, $B_{c_{1}}$ must be zero\footnote{After the completion of this work, a paper \cite{Dias:2013bwa} appeared where it is shown that indeed a $2+1$ dimensional holographic superconductor can behave as type I or II, depending on the value of the scalar field charge.}. From the holographic point of view, a 2+1 dimensional superconductor under a 3+1 electromagnetic field is to be type II because there is no way to exclude the magnetic field dynamically with the standard boundary conditions for the gauge field at the $AdS_{4}$ boundary, and therefore the magnetic field will be externally imposed.\footnote{I wish to thank the referee for this clarification.} The holographic approach discussed in this paper describes $3+1$ dimensional systems using $d=4+1$ $AdS$ spacetime models. In a $3+1$ dimensional system subjected to a $3+1$ electromagnetic field, both free energies scale with the volume (see Appendix), and hence there is a direct thermodynamical competition that can drive the system to a type I superconducting state.

\bigskip

This article is organized as follows. In section 2, we introduce a simple model and briefly review the holographic superconducting regime, without any magnetic field present. In section 3, a magnetic perturbation is turned on through the $U(1)$ gauge field of the bulk. Also, a small perturbation around the bulk scalar field condensed solution is turned on. We show that the system can be consistently described by a Ginzburg-Landau phenomenological description. Thereby we determine the Ginzburg-Landau parameter $\kappa$ for different values of the charge $q$ of the bulk scalar field in our model. We also calculate the free energy of the system using the Ginzburg-Landau approach and compare it near-$T_{c}$ with the free energy calculated through the standard holographic techniques. Finally, we compare the proposed Ginzburg-Landau approach with the methods developed in \cite{Herzog:2008he} for computing the Ginzburg-Landau theory parameters $\alpha$ and $\beta$. In section 4, we subject our system to a constant magnetic field solution. We calculate the critical magnetic field $B_{c}$ of the superconductor, and compare its near-$T_{c}$ behaviour with the results obtained in section 3.


\section{A minimal holographic superconductor in $d=4+1$ $AdS$}


\subsection{The model}

We will work using a minimal phenomenological model in $d=4+1$ $AdS$ spacetime, in the same spirit as in {\cite{Hartnoll:2008kx}}, containing a scalar field $\Psi$ and a $U(1)$ gauge field $A_{\mu}$
\begin{equation}
\label{basicmodel}
\mathcal{L} = R+\frac{12}{L^{2}}-\frac{1}{4}F^{\mu \nu}F_{\mu \nu}-\left|D\Psi \right|^{\,2}-M^{2}\left|\Psi\right|^{2}\,,
\end{equation}
where, $F_{\mu \nu}=\nabla_{\mu} A_{\nu}-\nabla_{\nu} A_{\mu}	$, and $D_{\mu}\Psi=\nabla_{\mu}\Psi-i q A_{\mu} \Psi$. The parameter $q$ corresponds to the charge of the scalar field and, as it will be shown below, different values of $q$ will correspond to superconducting systems with different critical temperature. The general equations of motion for this system are
\begin{eqnarray}
\label{generalpsieq}
D^{\,2}\Psi &=& M^{2}\Psi \,,\\
\label{generalgaugeeq}
\nabla_{\mu}F^{\mu\,\nu}&=&q J^{\nu}+q^{2}\left|\Psi\right|^{2}A^{\nu}\,,\\
R_{\mu\nu}-\frac{1}{2}g_{\mu\nu} \left(R + \frac{12}{L^2}\right)&=& \frac{1}{2}g_{\mu\nu}\left(-\frac{1}{4}F^2 -\left|D\Psi\right|^2 -M^{2}\left|\Psi\right|^{2}\right)\nonumber\\
\label{generaleinsteineq}
&&+\frac{1}{2}F^\lambda_{\;\;\mu} F_{\lambda \nu}+ D_{[\mu}\Psi D_{\nu]}\Psi\,,
\end{eqnarray}
where
\begin{equation}
J_{\mu}=i\left(\Psi^{*}\nabla_{\mu}\Psi - \Psi \nabla_{\mu} \Psi^{*}\right)\,.
\end{equation}

We will set $L=1$ for the rest of this paper. 
	
\subsection{The normal and superconducting phases}

In this section we will briefly review the normal and superconductor regimes of our model, with no external magnetic field to begin with, and with full backreaction included. As is usual, we use the following ansatz for the metric
\begin{equation}		
ds^{2} = -g(r) e^{-\chi(r)} dt^{2}+\frac{dr^{2}}{g(r)}+r^{2}\left( dx_{1}^{2} + dx_{2}^{2} + dx_{3}^{2} \right)\,,
\end{equation}
which is the most general ansatz with space-rotation and time-translation symmetry. We will demand that solutions for this ansatz are asymptotically $AdS$ and that they have a black hole geometry, with an outer event horizon at some $r=r_{h}$. For the scalar and gauge field we use the ansatz
\begin {equation}
A = \phi(r) dt\,,\;\;\;\;\;\;\;\;\; \Psi(r)= \frac{1}{\sqrt{2}}\psi(r)\,,
\end{equation}
where $\psi$ is a real function. Introducing the new coordinate $z=r_{h}/r$, equations (\ref{generalpsieq}-\ref{generaleinsteineq}) under this ansatz turn to be
\begin{eqnarray}
\label{psieqz}
\psi''+\left(  -\frac{\chi'}{2}-\frac{1}{z}+\frac{g'}{g} \right) \psi' +\frac{r_{h}^{2}}{z^{4} }\left(  \frac{e^{\chi} q^{2} \phi^{2}}{g^{2}} - \frac{M^{2}}{g} \right) \psi &=& 0 \,,\\
\label{phieqz}
\phi'' + \left(  \frac{\chi'}{2}-\frac{1}{z} \right) \phi' - \frac{r_{h}^{2} q^{2} \psi^{2}}{z^{4} g} \phi &=&0\,,\\
\label{chieqz}
3\chi' - z \psi'^{2} -\frac{e^{\chi} q^{2} \phi^{2} \psi^{2}}{z^{3} g^{2}} &=& 0 \,,\\
\label{geqz}
\frac{1}{2} \psi'^{2} +\frac{e^{\chi} \phi'^{2}}{2 g} - \frac{3 g'}{z g} + \frac{6}{z^{2}} -\frac{12 r_{h}^{2}}{z^{4} g} + \frac{r_{h}^{2} M^{2} \psi^{2}}{2 z^{4} g} + \frac{e^{\chi} r_{h}^{2} q^{2} \phi^{2} \psi^{2}}{2 z^{4} g^{2}} &=& 0 \,.
\end{eqnarray}

This system of equations admit a $\psi(z) = 0$ solution. This no-hair solution is given by
\begin{eqnarray}
g(r) &=&\frac{r_{h}^{2}}{z^2}+\frac{z^4 \rho ^2}{3 r_{h}^4}-\frac{z^2 \left(3 r_{h}^6+\rho ^2\right)}{3 r_{h}^4}\,,\\
\chi(r) &=& 0\,,\\
\phi(r) &=& \frac{\rho}{r_{h}^{2}} \left(1-z^{2}\right) \,,
\end{eqnarray}
which is the usual Reissner-Nordstr\"{o}m-$AdS$ solution, and corresponds to the normal phase of the superconductor. 

\bigskip

We will now consider solutions with scalar hair $\psi\neq 0$. We will set $M^{2}L^{2}=-3$ for the scalar field mass, which is above the Breitenlohner-Freedman bound $M^{2}_{\text{BF}}L^{2}=-4$. This choice of mass appears naturally in top-down models of holographic superconductors coming from consistent truncations of supergravity \cite{Gubser:2009qm, Aprile:2011uq}\footnote{These models have a different potential from ours, arising from higher order terms in $\psi$. However, they have the same critical temperature, since this only depends on the values of $m$ and $q$. }. With this choice,  $\psi$ behaves at $z \rightarrow \infty$ as 
\begin{equation}
\label{assymptpsi}
\psi \approx \mathcal{O}_{1} \frac{z}{r_{h}} + \mathcal{O}_{3}\frac{z^{3}}{r_{h}^{3}} + \ldots
\end{equation}
while for the gauge field the near-boundary behaviour is
\begin{equation}
\label{assymptphi}
\phi \approx\mu- \rho\frac{z^{2}}{r_{h}^{2}} + \ldots
\end{equation}

According to the gauge-gravity correspondence, $\mathcal{O}_{3}$ corresponds to the vacuum expectation value of an operator of dimension $3$ in the dual field theory, while $\mathcal{O}_{1}$ corresponds to a source to that same operator. Also, $\mu$ and $\rho$ will correspond to the chemical potential and charge density of the dual field theory, respectively. To solve our equations of motion, we will impose the boundary condition $\mathcal{O}_{1}=0$ in (\ref{assymptpsi}) and take $\mathcal{O}_{3}$ as the superconductor order parameter. Setting the source to zero will result in spontaneous breaking of the global $U(1)$ symmetry in the dual field theory and the system enters then in a superconducting phase \cite{Hartnoll:2008kx, Gubser:2008px}. 

\bigskip

We will choose to work in the canonical ensemble, fixing $\rho=1$. As mentioned above, we will also impose $g(z=1)=0$ for some non-zero value of $r_{h}$ in order to have black hole solutions to our ansatz and introduce temperature to the dual field theory. The Hawking temperature of the system will be given by
\begin{equation}
T_{H}=-\frac{e^{\chi} g'}{4\pi r_{h}}\Bigg|_{z=1}\,.
\end{equation} 
From equation (\ref{psieqz}) for $\psi$ we see that regularity of the solutions at the horizon $z=1$ requires that
\begin{equation}
\psi'(1) = \frac{r_{h}^{2} M^{2} \psi(1)}{g'(1)}\,. 
\end{equation}
Regularity at the horizon also requires $\phi(1)=0$. The model has the following scaling symmetries
\begin{eqnarray}
\label{scaleinvariance1}
e^{\chi} \rightarrow\ a^{2} e^{\chi}\,,\;\;t \rightarrow a t\,,\;\; \phi \rightarrow \phi /a\,,\\
\label{scaleinvariance2}
r \rightarrow a r\,, \;\; \left( t, x_{i} \right) \rightarrow \left( t, x_{i} \right) / a\,\;\; g \rightarrow a^{2} g\,,\;\; \phi \rightarrow a \phi\,.
\end{eqnarray}
This scale invariance helps us to further reduce the number of independent parameters in our model to only one, which we will take to be the temperature of the black hole. Solutions to equations (\ref{psieqz})-(\ref{geqz}) are found via the shooting method, enforcing the no-source condition mentioned above for $\psi$.

\begin{figure}[t!]
\begin{center}
\begin{picture}(250,150)
\put(0,0){\includegraphics*[width=3.0in,angle=0]{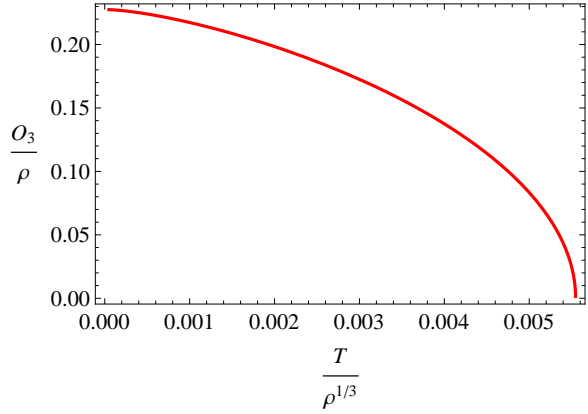}}%
\end{picture}
\caption{\label{cond} The value of the condensate as a function of temperature, for $q=1$. In this case, $T_{c}=0.0055$ approximately.}
\end{center}
\end{figure}

\begin{figure}[t!]
\begin{center}
\begin{picture}(250,160)
\put(0,0){\includegraphics*[width=.6\linewidth,angle=0]{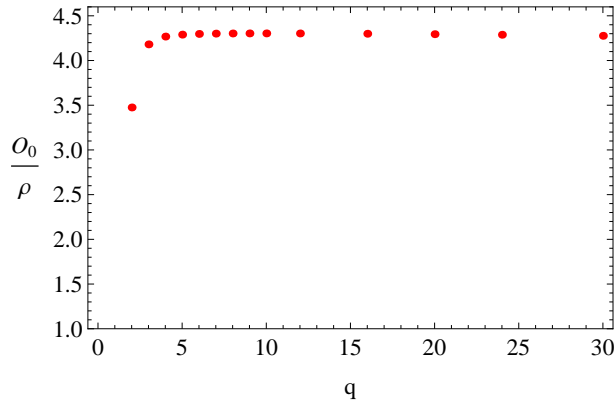}}%
\end{picture}
\caption{\label{O0fig} The value of the near-$T_{c}$ coefficient $\mathcal{O}_{0}$ (see eq. (\ref{O0def})) as a function of the scalar field charge $q$.}
\end{center}
\end{figure}

\begin{figure}[t!]
\begin{center}
\begin{picture}(250,160)
\put(0,0){\includegraphics*[width=.6\linewidth,angle=0]{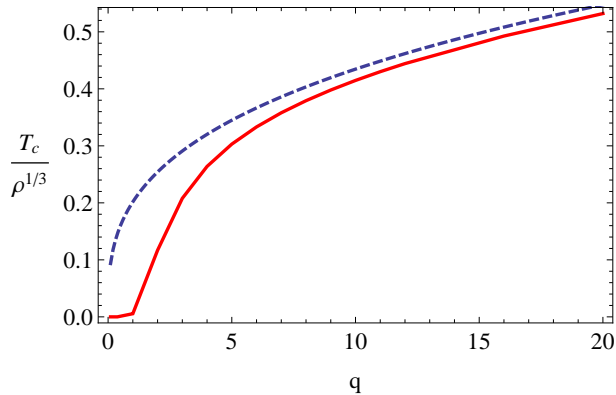}}%
\end{picture}
\caption{\label{Tccomparisons} The solid line represents the value of the critical temperature $T_{c}$ as a function of the charge $q$. The dashed line represents the analytical approximation (\ref{matchingT}).}
\end{center}
\end{figure}

\bigskip

In figure (\ref{cond}) we show the behaviour of the order parameter $\mathcal{O}_{3}$ as a function of temperature for the case $q=1$, signaling condensation below some critical temperature $T_{c}$. One can find by a numerical analysis for different values of $q$ that near $T_{c}$ the condensate behaves as
\begin{equation}
\label{O0def}
\mathcal{O}_{3}\sim\mathcal{O}_{0}\left(1-T/T_{c}\right)^{1/2}\,.
\end{equation}
The behaviour of the coefficient $\mathcal{O}_{0}$ as a function of the scalar field charge $q$ is shown in figure (\ref{O0fig}). For large values of $q$, we find that $\mathcal{O}_{0}\sim \text{const.}$

\bigskip

In the bold line of figure (\ref{Tccomparisons}) we show how the critical temperature $T_{c}$ behaves for different values of the charge $q$. As in the $2+1$ dimensional case of \cite{Hartnoll:2008kx}, the behaviour of $T_{c}$ near zero $q$ is caused because the charged scalar field backreacts to the metric more strongly in that region, decreasing the temperature. Since we have a one-to-one relation between $T_{c}$ and $q$, we will use $q$ to vary the critical temperature of our model. Therefore, we will have a set of different superconducting systems characterized by different $q$. 

For large values of $q$ one can obtain a fair analytical approximation for $T_{c}$ using the matching method introduced in \cite{Gregory:2009fj}, getting
\begin{equation}
\label{matchingT}
T_{c}^{\text{large $q$}} = \frac{1}{\pi} \left(\sqrt{\frac{5}{309}} 2\rho\,q\right)^{\frac{1}{3}}\,.
\end{equation}
This is shown as a dashed line in figure  (\ref{Tccomparisons}).\footnote{For applications of the matching method on the study of magnetic effects in holographic superconductors, see. e.g. \cite{Ge:2011cw, Ge:2012vp}.}
 
\pagebreak
\section{Ginzburg-Landau description of the holographic superconductor.}

We introduce our Ginzburg-Landau interpretation of the dual field theory by first studying the system under a small perturbation of the gauge field on the bulk.

\subsection{A magnetic perturbation}

We now add a small magnetic perturbation of the gauge field, in the specific form
\begin{equation}
\label{Aperturb}
A= \phi(r)\,dt + \delta A_{x}(r,t,y)\,dx;
\end{equation}
with
\begin{equation}
\delta A_{x}(t,r,y) = e^{-i\,\omega\,t + i\,k\,y}A_{x}(r)\,,\;\;\;\;\;\;\; \left|A_{x}\right| \ll 1
\end{equation}
This perturbation has an harmonic dependence on time and carries momentum along the $y$-direction. To linearized level, the equation of motion for $A_{x}$ in the $z$ coordinate is given by
\begin{equation}
\label{ax_EOM_z}
A_{x}''+\left(\frac{g'}{g}+\frac{1}{z}-\frac{\chi'}{2}\right)\,A_{x}' +\frac{r_{h}^{2}}{z^{2}g}\left(\frac{e^{\chi}\omega^{2}}{z^{2}g}-\frac{k^{2}}{r_{h}^{2}}-\frac{q^{2}\psi^{2}}{z^{2}}\right)\,A_{x} = 0\,.
\end{equation}
We will work in the low-frequency/small-momentum regime, where $k\,,\omega\,$ are much smaller than the scale of the condensate, so that quadratic terms in $k$, $\omega$ can be neglected in (\ref{ax_EOM_z}). To solve this equation, we use the following boundary conditions
\begin{equation}
A_{x}(1)=A_{0}\,,\;\;\;\;\;\;\;A'_{x}(1)= -\frac{6 q^2 r_{h}^2 \psi_{0}^2}{e^{\chi0} \phi_{0}^2+r_{h}^2 \left(M^2 \psi_{0}^2-24\right)}\,A_{0}\,,
\end{equation}
where we use the notation $\psi_{0}=\psi(1)$, $\phi'_{0}=\phi(1)$, and where the second condition is needed for regularity at the horizon. As before, $M^{2}=-3$. Since the equation (\ref{ax_EOM_z}) is linear, with no loss of generality we set $A_{0}=1$. 

From equation (\ref{ax_EOM_z}) we can read the behaviour of $A_{x}$ at $z\rightarrow 1$
\begin{equation}
\label{assymptAx}
A_{x}=A_{x}^{(0)} + J_{x} \frac{z^{2}}{r_{h}^{2}} + \ldots\,.
\end{equation}

\begin{figure}[t!]
\begin{center}
\begin{picture}(250,170)
\put(0,0){\includegraphics*[width=.6\linewidth,angle=0]{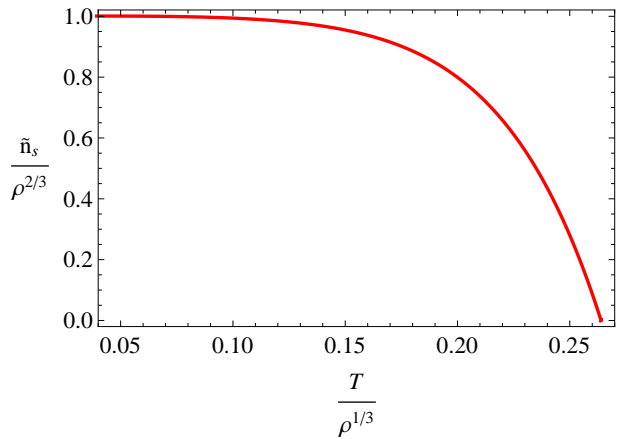}}%
\end{picture}
\caption{\label{ns4} Value of $\tilde{n}=\frac{q^{2}}{m}n_{s}$ as a function of temperature, for the $q=4$ case. }
\end{center}
\end{figure}

According to the $AdS/CFT$ dictionary, $A_{x}^{(0)}$ and $J_{x}$ correspond to a vector potential and the conjugated current on the dual field theory, respectively. We can identify these asymptotic values with the London current on the dual superconducting field theory (see eq. (\ref{londoneqns}))
\begin{equation}
\label{HolographicLondonEquation}
J_{x} = -\frac{q^{2}}{m}n_{s} A_{x}^{(0)}\,,
\end{equation}
were $n_{s}$ is the number density of superconducting carriers and $q$ and $m$ are the charge and mass of the superconducting carriers, respectively. At this point, it is worth mentioning that, as stated in \cite{Hartnoll:2008kx}, the London equation is valid only when $k$ and $\omega$ are small compared to the scale of the condensate, in consistency with our low-frequency/small-momentum regime. From (\ref{HolographicLondonEquation}) we can read the value of the quantity $q^{2} n_{s}/m$ holographically as
\begin{equation}
\label{holographicns}
\frac{q^{2}}{m}n_{s}=-\frac{J_{x}}{A_{x}^{(0)}}\,.
\end{equation}
For simplicity, we define the quantity
\begin{equation}
\label{nstilde}
\tilde{n}_{s}\equiv \frac{q^{2}}{m}n_{s}\,,
\end{equation}
which is a rescaling of the carrier number density. Numerically one finds that $\tilde{n}_{s}$ behaves near $T_{c}$ as $\tilde{n}_{s}\sim \left(1-T/T_{c}\right)$. The value of $\tilde{n}_{s}$ as a function of temperature for charge $q=4$ is shown in figure (\ref{ns4}).

\pagebreak
\subsection{Ginzburg-Landau interpretation of the dual field theory}

In this section we will implement a phenomenological Ginzburg-Landau description of our superconducting system by assuming that the dual $d=3+1$ field theory at non-zero temperature can be described phenomenologically by an effective Ginzburg-Landau field theory. This will be given by a vector field $\mathcal{A}_{\mu}$, $\mu=0\,,\ldots\,, 3$, and a scalar field $\Psi_{\text{GL}}$ which acts as an order parameter for the theory and effectively represents the operator that condenses in the underlying dual field theory, which in principle could have very different degrees of freedom. This Ginzburg-Landau description is only valid near the critical temperature, where the order parameter $\Psi_{\text{GL}}$ is small, and where the effective action for the dual field theory can be written as
\begin{equation}
\label{effectiveS}
S_{\text{eff}}\approx \frac{1}{T}\int d^{3}x\left\{\alpha\left|\Psi_{\text{GL}}\right|^{2}+\frac{\beta}{2}\left|\Psi_{\text{GL}}\right|^{4}+\frac{1}{2 m}\left|D_{i} \Psi_{\text{GL}}\right|^{2}+\ldots\right\}\,,
\end{equation}
where $D_{i}=\partial_{i}-iq\,\mathcal{A}_{i}$, and $\alpha$ and $\beta$ are phenomenological parameters with a temperature dependence\footnote{ For a discussion about effective field approximations in the dual field theory, see \cite{Domenech:2010nf}. For other works on aspects of Ginzburg-Landau theory in the context of holography, see, e.g. \cite{Herzog:2008he, Maeda:2008ir, Yin:2013fwa, Banerjee:2013maa}.}. According to the $AdS$/$CFT$ dictionary, the vector components $\mathcal{A}_{0}$ and $\mathcal{A}_{x}$ correspond respectively to the chemical potential $\mu$ in (\ref{assymptphi}) and to $A_{x}^{(0)}$ in (\ref{assymptAx}). We have consistently identified the charge of the superconducting carrier of the phenomenological Ginzburg-Landau Lagrangian with the charge of the bulk scalar field $q$. We will be mainly interested in electromagnetic phenomena present in superconductivity, which require a dynamical gauge field in the boundary theory. However, we know that the  $U(1)$ local symmetry  in the bulk translates to a global $U(1)$ symmetry in the boundary according to the gauge/gravity dictionary. In order to overcome this, we will assume that the $U(1)$ global symmetry in the boundary can be promoted to local, by adding a $F^{2}$ term using the procedure described in \cite{Hartnoll:2008kx}. Indeed, this is the underlying procedure behind most studies of magnetic phenomena in holographic superconductivity. In terms of our current effective field theory description of the boundary theory, this will mean that the Ginzburg-Landau theory approach to electromagnetic phenomena can be applied in our case, especially concerning its determination of the critical magnetic field and of the Ginzburg-Landau parameter $\kappa$, which requires a balance between the superconducting and the purely magnetic parts of the free energy of the system (see Appendix).

\bigskip

The VEV of the scalar operator that condenses in the underlying dual field theory will be proportional to $\mathcal{O}_{3}$ to the required power to match dimensions. The Ginzburg-Landau order parameter $\Psi_{\text{GL}}$ has mean field critical exponent $1/2$. Then, in order to match this critical exponent with the critical exponent of $\mathcal{O}_{3}$ we must identify
\begin{equation}
\label{id}
\left|\Psi_{\text{GL}}\right|^{2}=N_{q}\mathcal{O}_{3}^{2}\,,
\end{equation}
where $N_{q}$ is a proportionality constant that depends on the value of the charge $q$ of the scalar bulk field.

\begin{figure}[t!]
\begin{center}
\begin{picture}(250,170)
\put(0,0){\includegraphics*[width=.6\linewidth,angle=0]{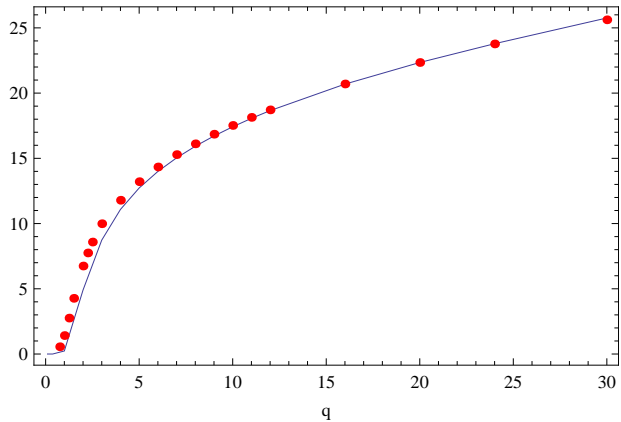}}%
\end{picture}
\caption{\label{numeqg} Comparison between $q\,\mathcal{O}_{3}^{2}/\tilde{n}_{s}$, corresponding to red points, and $C_{0} T_{c}(q)$, corresponding to continuous line. }
\end{center}
\end{figure}

\bigskip

Regarding the parameters $\alpha$, $\beta$ shown in (\ref{effectiveS}), one sets $\beta > 0$ in order for the lowest free energy to be at finite $\left|\Psi_{\text{GL}}\right|^{2}$. Also, in order to have a superconducting phase, one requires that $\alpha<0$. All definitions and conventions that will be used regarding the Ginzburg-Landau theory can be found in the Appendix, where we have set the physical constants $\hbar=1$ and $\mu_{0}=4\pi$ (their values in natural units), while preserving numerical factors. The superconducting carrier mass $m$ can be absorbed into a redefinition of the other parameters, so, with no loss of generality, we will set $m=1$. 

\bigskip

\begin{figure}[t!]
\begin{center}
\begin{picture}(250,170)
\put(0,0){\includegraphics*[width=.6\linewidth,angle=0]{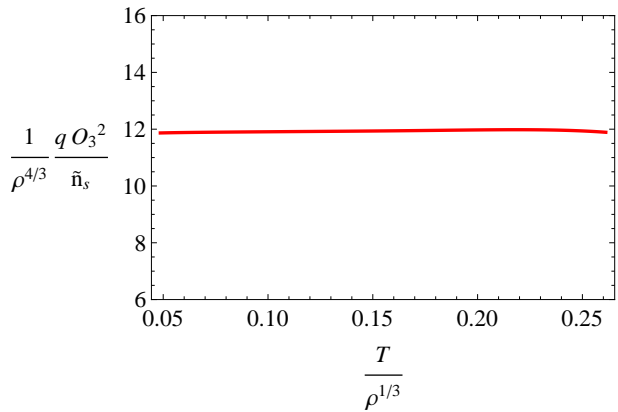}}%
\end{picture}
\caption{\label{ratio2} Value of the ratio $q\frac{\mathcal{O}_{3}^{2}}{\tilde{n}_{s}}$ as a function of temperature, for the case $q=4$. }
\end{center}
\end{figure}

At this point, we have two phenomenological Ginzburg-Landau parameters $\alpha$ and $\beta$, and introduced the proportionality constant $N_{q}$. We should be able to fully determine them in order for our Ginzburg-Landau description to be as complete and consistent as possible. In order to do it, we will make use of the numeric identity
\begin{equation}
\label{numeq}
q\frac{\mathcal{O}_{3}^{2}}{\tilde{n}_{s}}=C_{0} T_{c}(q)\,,
\end{equation}
where the ratio at the left hand side is evaluated at the critical temperature, and $C_{0}$ is a proportionality constant, approximately equal to $C_{0}\approx 41.99$. In figure (\ref{numeqg}) we show how this equality holds for various values of $q$. To have a better understanding of this equality, one can see through the matching method in the large $q$ limit that $\mathcal{O}_{3}\sim \frac{T_{c}^{3}}{q} \left(1-T/T_{c}\right)^{1/2}$ and $\tilde{n}_{s} \sim T_{c}^{2}\left(1-T/T_{c}\right)$, so the left hand side of the equality goes as $q\,\mathcal{O}_{3}^{2}/\tilde{n}_{s} \sim T_{c}^{4}/q$, and because $q \sim T_{c}^{3}$ (see (\ref{matchingT})), we indeed have $q\,\mathcal{O}_{3}^{2}/\tilde{n}_{s} \sim T_{c}$. Another point worth mentioning is that the left hand side of equation (\ref{numeq}) is constant as a function of temperature, for most values of $q$. This is shown in figure (\ref{ratio2}), where we plot $q\,\mathcal{O}_{3}^{2}/\tilde{n}_{s}$ versus temperature, for the $q=4$ case.

\bigskip

\begin{figure}[t!]
\begin{center}
\begin{picture}(250,170)
\put(0,0){\includegraphics*[width=.6\linewidth,angle=0]{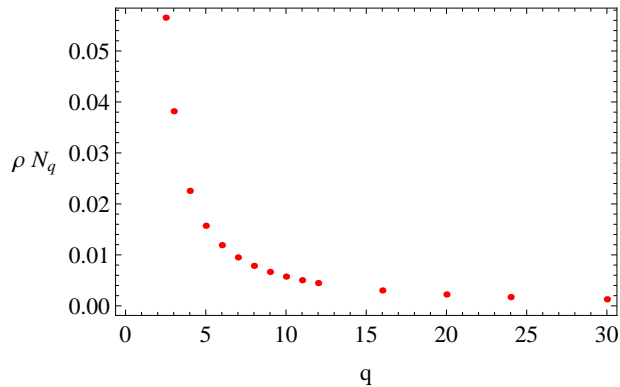}}%
\end{picture}
\caption{\label{Nqfig} Value of the proportionality factor $N_{q}$ as a function of the scalar field charge $q$. }
\end{center}
\end{figure}

Rewriting (\ref{numeq}) in terms of $n_{s}$ instead of $\tilde{n}_{s}$, we have
\begin{equation}
\label{numeq2}
\frac{\mathcal{O}^{2}_{3}}{q\, n_{s}}=C_{0}T_{c}\,.
\end{equation}
According to the Ginzburg-Landau theory, the relation between the order parameter $\left|\Psi_{\text{GL}}\right|$ and the charge carrier density $n_{s}$ is given by (see (\ref{psins}))
\begin{equation}
\label{holographicpsins}
\left|\Psi_{\text{GL}}\right|^{2}= n_{s}\,.
\end{equation}
Substituting our identification (\ref{id}) in (\ref{holographicpsins}), and matching with (\ref{numeq2}) we obtain
\begin{equation}
\label{Nq}
N_{q}=\frac{1}{q\,C_{0} T_{c}(q)}\,.
\end{equation}
The behaviour of $N_{q}$ as a function of $q$ is shown in figure (\ref{Nqfig}). For large $q$ we find $N_{q}\sim q^{-4/3}$.

\bigskip

In order to determine the remaining parameters, we must calculate first the Ginzburg-Landau coherence length $\xi$. To do this, we consider small fluctuations around the condensed phase of our system in the bulk. More concretely, we write the original complex scalar field $\Psi$ in our model (\ref{basicmodel}) as

\begin{equation}
\Psi(r,y) = \frac{1}{\sqrt{2}}\left(\psi(r)+e^{i\,k\,y}\eta(r)\right)\,,
\end{equation}
where $\psi$ is the full back-reacted solution associated with the order parameter $\mathcal{O}_{3}$ described in section 2, and the term $e^{i\,k\,y}\eta(r)$ is a small fluctuation ($\left|\eta\right|\ll 1$) around this condensed solution. The equation of motion for $\eta$ to linearized level is
\begin{equation}
\label{etaequ}
\eta''+\left(\frac{g'}{g}-\frac{\chi'}{2}-\frac{1}{z}\right)\eta'+\frac{1}{z^{2}g}\left(\frac{e^{\chi}q^{2}r_{h}^{2}\Phi^{2}}{z^{2}g}-\frac{M^{2}r_{h}^{2}}{z^{2}}-k^{2}\right)\eta=0\,,
\end{equation}
which can be put as in the form of an eigenvalue equation
\begin{equation}
\label{eigeneta}
\mathcal{L}\left\{\eta\right\}=k^{2}\eta\,,
\end{equation}
with $\mathcal{L}$ the same linear operator that acts on $\psi$. The boundary conditions at the horizon $z=1$ are:
\begin{equation}
\label{horizoneta}
\eta(1)=\eta_{0}\,,\hspace{40pt} \eta'(1)=-\frac{6\left(k^{2}+M^{2}r_{h}^{2}\right)}{e^{\chi_{0}}\Phi_{0}^{2}+r_{h}^{2}\left(M^{2}\psi_{0}^{2}-24\right)}\eta_{0}\,,
\end{equation}
while near $z=0$ we will have the asymptotic behaviour
\begin{equation}
\eta(z) \approx \left(\delta\mathcal{O}_{1}\right)\frac{z}{r_{h}} + \left(\delta\mathcal{O}_{3}\right) \frac{z^{3}}{r_{h^{3}}}+\cdots\,,
\end{equation}
and will demand the same conditions as for $\psi$, namely $\left(\delta\mathcal{O}_{1}\right)=0$. Since, as will be seen below, we will not be concerned with the absolute normalization of $\eta$, we will take advantage of the linearity of (\ref{etaequ}) and set $\eta_{0}=1$.

\begin{figure}[t!]
\begin{center}
\begin{picture}(250,170)
\put(0,0){\includegraphics*[width=.6\linewidth,angle=0]{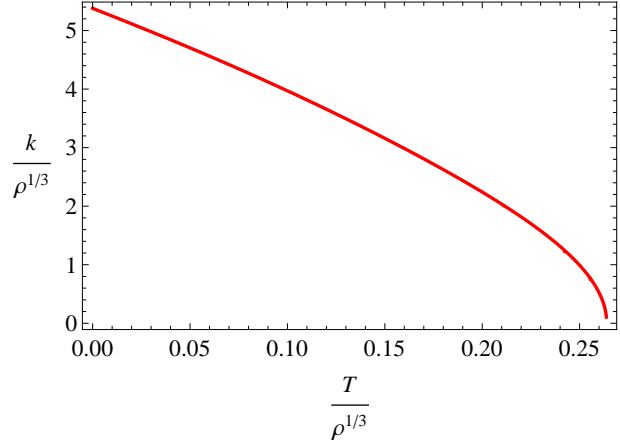}}%
\end{picture}
\caption{\label{kg} Value of the wave number $k$ as a function of temperature, for the case $q=4$. }
\end{center}
\end{figure}

\begin{figure}[t!]
\begin{center}
\begin{picture}(250,170)
\put(0,0){\includegraphics*[width=.6\linewidth,angle=0]{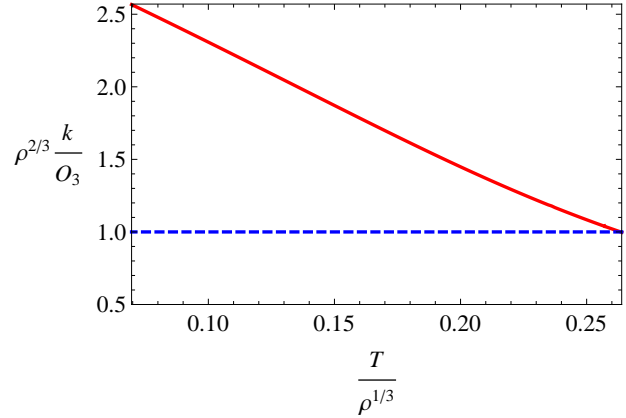}}%
\end{picture}
\caption{\label{fitg} Value of the ratio $k/\mathcal{O}_{3}$ as a function of temperature, for the case $q=4$. The dashed line corresponds to the respective value of $A_{q}$, which in this case is close to one.  }
\end{center}
\end{figure}

\bigskip

\pagebreak

\begin{figure}
\centering
\begin{subfigure}{.5\textwidth}
  \centering
 \begin{picture}(250,150)
\put(0,0){\includegraphics*[width=1\linewidth]{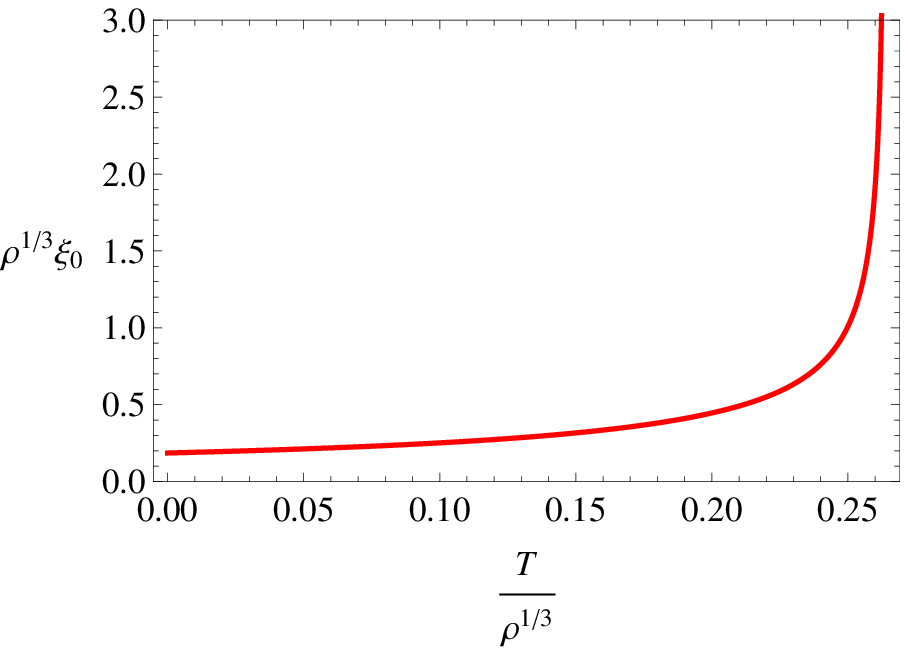}}%
\end{picture}
  \caption{Coherence length $\xi_{0}$, $q=4$}
  \label{xi0g}
\end{subfigure}%
\begin{subfigure}{.5\textwidth}
  \centering
	\begin{picture}(250,150)
\put(10,0){\includegraphics*[width=1\linewidth]{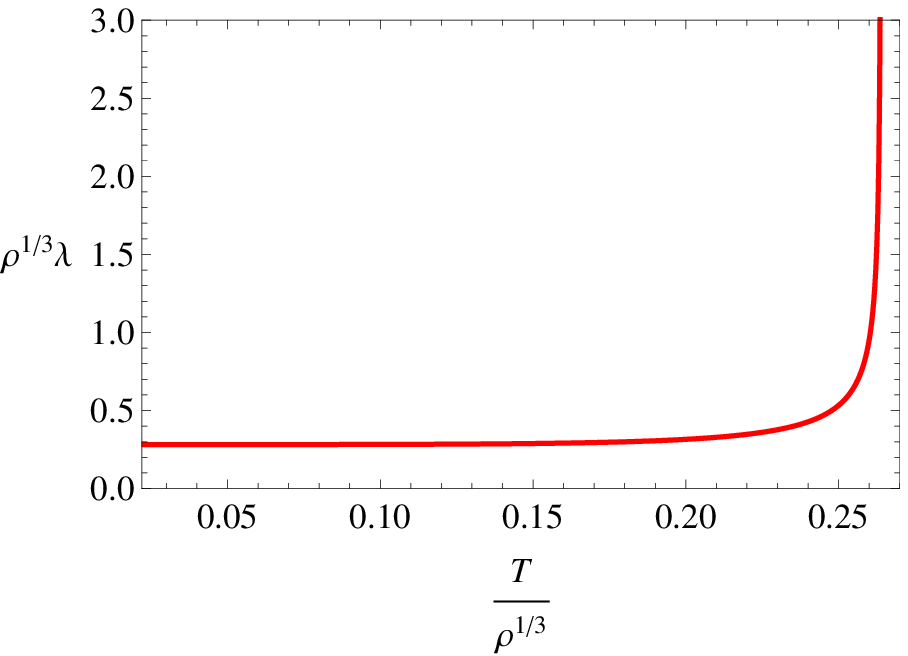}}%
\end{picture}
  \caption{Penetration length $\lambda$, $q=4$}
  \label{lambda4}
\end{subfigure}
\caption{Value of the characteristic lengths $\xi_{0}$ and $\lambda$ as a function of temperature, for the case $q=4$.}
\end{figure}

As noted in \cite{Maeda:2008ir}, the coherence length of the superconducting system is equal to the correlation length $\xi_{0}$ of the order parameter. In turn, the correlation length is the inverse of the pole of the correlation function of the order parameter written in Fourier space
\begin{equation}
\label{correlation}
\left\langle \mathcal{O}(k)\mathcal{O}(-k)\right\rangle \sim \frac{1}{\left|k\right|^{2}+1/\xi_{0}^{2}}\,.
\end{equation}
This pole will be given by the eigenvalue of (\ref{eigeneta}). Therefore,  we must solve equation (\ref{etaequ}) and calculate the value of the wave number $k$ consistent with the desired boundary conditions for $\eta$. This was done near the critical temperature. The behaviour of the wave number $k$ as a function of temperature is shown in figure (\ref{kg}), for $q=4$. From the wave number $k$ we obtain the coherence length $\xi_{0}$ simply as 
\begin{equation}
\label{xik}
\left|\xi_{0}\right|=\frac{1}{\left|k\right|}\,.
\end{equation}
whose behaviour as a function of temperature is shown in figure (\ref{xi0g}), also for the value $q=4$. 

\bigskip

\begin{figure}[t!]
\begin{center}
\begin{picture}(250,170)
\put(0,0){\includegraphics*[width=.6\linewidth,angle=0]{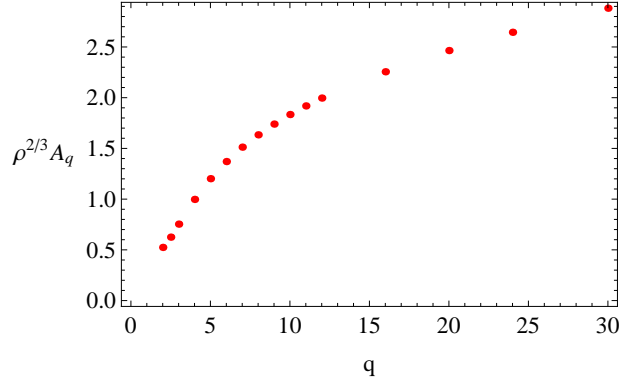}}%
\end{picture}
\caption{\label{Aqg} Value of the proportionality constant $A_{q}$ as a function of the scalar field charge $q$.  }
\end{center}
\end{figure}

It should be pointed out that the wave number $k$ near the critical temperature becomes equal to the order parameter $\mathcal{O}_{3}$ times a proportionality constant $A_{q}$, which depend on the value of the charge $q$ considered. The value of $A_{q}$ is given by the ratio between $k$ and $\mathcal{O}_{3}$ evaluated at $T_{c}$
\begin{equation}
\label{Aqdef}
A_{q}=\frac{k}{\mathcal{O}_{3}}\bigg|_{T=T_{c}}\,,
\end{equation}
which, for every case considered, was a finite number. The value of the ratio $k/\mathcal{O}_{3}$ as a function of temperature can be seen in figure (\ref{fitg}), for $q=4$. The value of $A_{q}$ as a function of the charge $q$ is shown in figure (\ref{Aqg}), and is found numerically to behave as $q^{1/3}$ for large values of $q\,$. From (\ref{xik}) and (\ref{Aqdef}), one has near the critical temperature
\begin{equation}
\label{xiTc}
\frac{1}{\xi_{0}}\approx A_{q} \mathcal{O}_{3}\,,\hspace{30pt} (T\approx T_{c})\,.
\end{equation}

\bigskip

\begin{figure}
\centering
\begin{subfigure}{.5\textwidth}
  \centering
 \begin{picture}(250,150)
\put(0,0){\includegraphics*[width=1\linewidth]{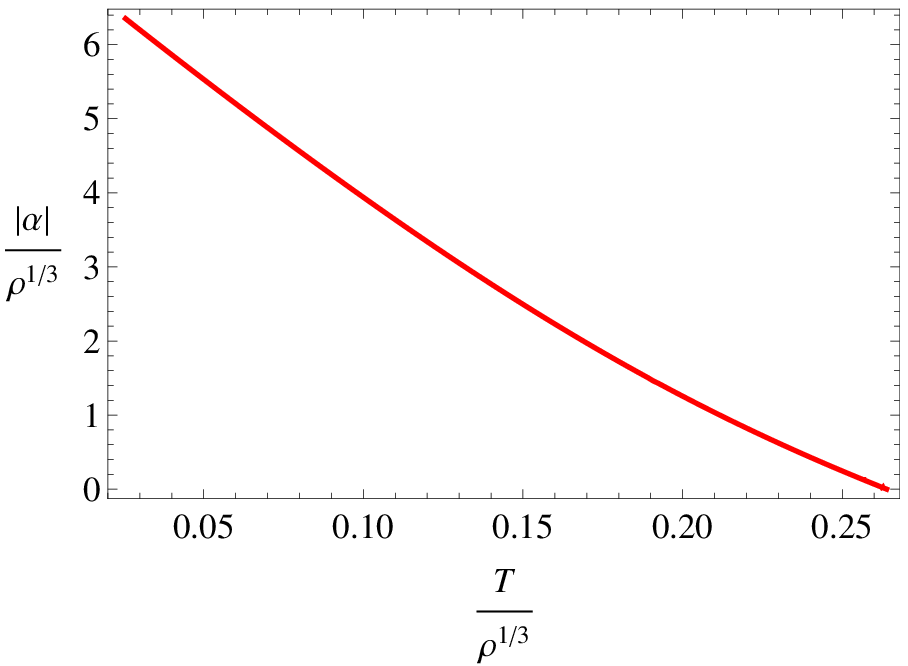}}%
\end{picture}
  \caption{$\alpha$}
  \label{alphag}
\end{subfigure}%
\begin{subfigure}{.5\textwidth}
  \centering
	\begin{picture}(250,150)
\put(10,0){\includegraphics*[width=1\linewidth]{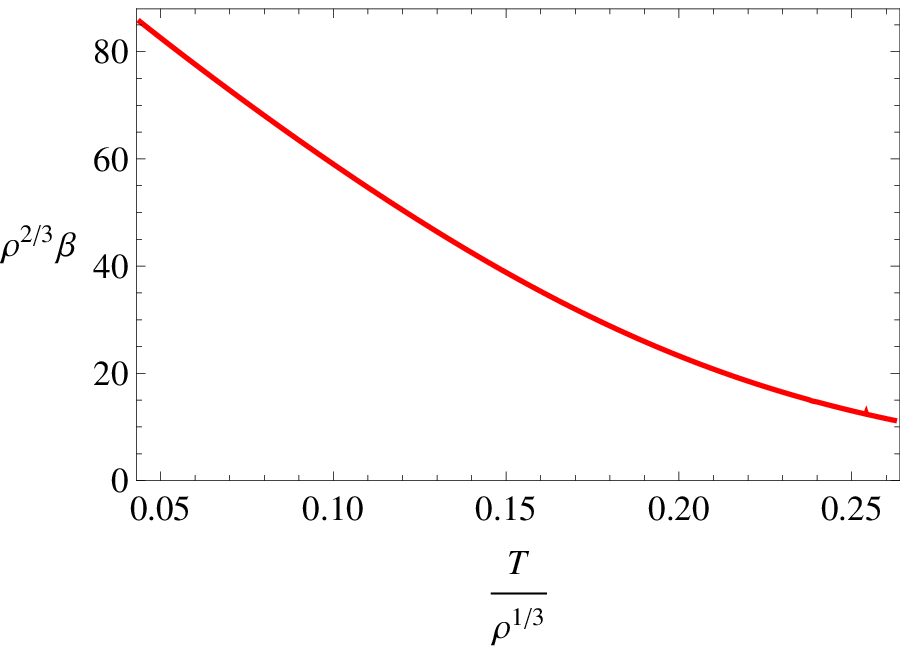}}%
\end{picture}
  \caption{$\beta$}
  \label{betag2}
\end{subfigure}
\caption{Value of the Ginzburg-Landau parameters $\alpha$ and $\beta$ as a function of temperature, for the case $q=4$.}
\end{figure}

With the calculation of the correlation length of the order parameter, and its identification as the superconductor coherence length, we now resort to the Ginzburg-Landau theory relation (\ref{correl}), which gives us the parameter $\left|\alpha\right|$ as
\begin{equation}
\label{alpha}
\left|\alpha\right|=\frac{1}{4\,\xi_{0}^{2}}\,.
\end{equation}
Since, as we mentioned above, near the critical temperature $\xi_{0} \approx A_{q}/\mathcal{O}_{3}$, then
\begin{equation}
\label{alphaTc}
\left|\alpha\right|\approx\frac{A_{q}^{2}}{4}\,\mathcal{O}_{3}^{2}\sim \left(1-T/T_{c}\right)\,,\hspace{40pt}(T\approx T_{c})
\end{equation}
which is the correct near-critical temperature behaviour for $\left|\alpha\right|$ according to Ginzburg-Landau theory. In figure (\ref{alphag}), we show the behaviour of $\alpha$ as a function of temperature, for the case $q=4$.

\bigskip

\pagebreak

In order to calculate the remaining Ginzburg-Landau parameter $\beta$, we will assume that the superconducting order parameter $\left|\Psi_{\text{GL}}\right|$ does not differ significantly from (see (\ref{PsiMinimum}))
\begin{equation}
\label{psideep}
\left|\Psi_{\infty}\right|^{2}=\frac{\left|\alpha\right|}{\beta}\,,
\end{equation}
which is the value of the order parameter that minimizes the Ginzburg-Landau free energy and physically is the value of $\left|\Psi_{\text{GL}}\right|$ deep inside the volume of the superconductor. As stated in the Appendix, this can only be so in the case where the external fields and gradients are negligible. This is indeed the case for our gauge perturbation (\ref{Aperturb}). Substituting our identification (\ref{id}) in (\ref{psideep}) we get
\begin{equation}
N_{q}\mathcal{O}_{3}^{2}=\frac{\left|\alpha\right|}{\beta}\,,
\end{equation}
from where we obtain, making use of (\ref{Nq}) and (\ref{alpha})
\begin{equation}
\label{beta1}
\beta=\frac{q\,C_{0}T_{c}(q)}{4 }\frac{1}{\xi_{0}^{2}\mathcal{O}_{3}^{2}}\,.
\end{equation}
In figure (\ref{betag2}) we show the behaviour of $\beta$ as a function of temperature, for the $q=4$ case.

\bigskip

\begin{figure}
\centering
\begin{subfigure}{.5\textwidth}
  \centering
 \begin{picture}(250,150)
\put(0,0){\includegraphics*[width=1\linewidth]{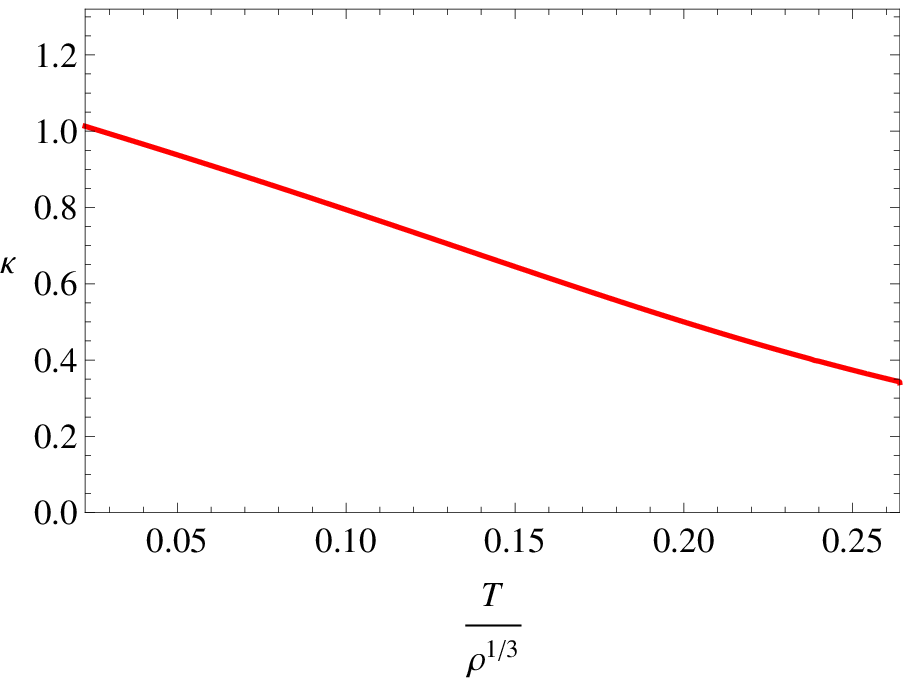}}%
\end{picture}
  \caption{$q=4$}
  \label{kg4}
\end{subfigure}%
\begin{subfigure}{.5\textwidth}
  \centering
	\begin{picture}(250,150)
\put(10,0){\includegraphics*[width=1\linewidth]{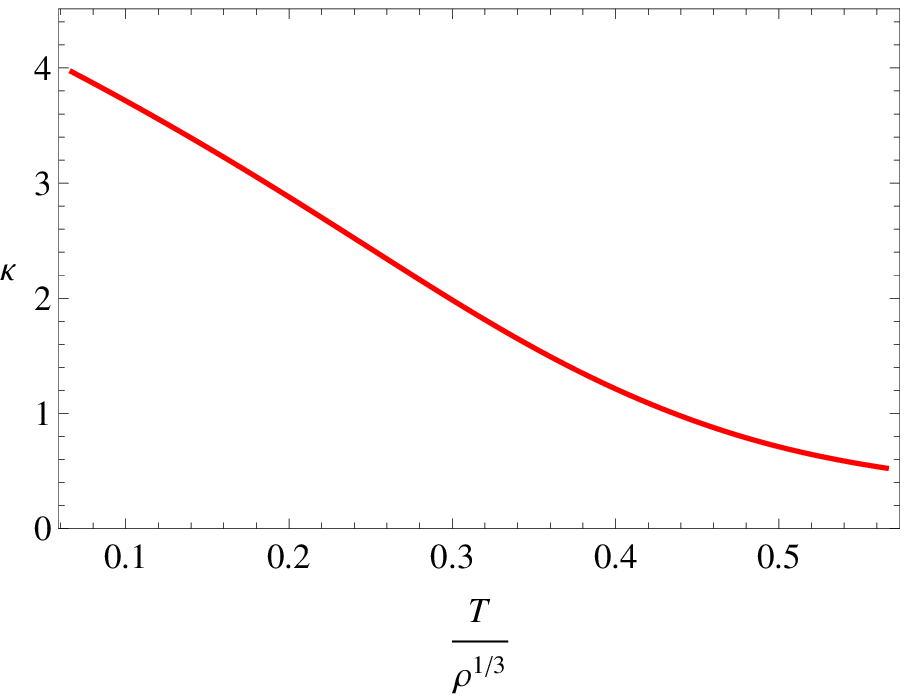}}%
\end{picture}
  \caption{$q=24$}
  \label{kg24}
\end{subfigure}
\caption{Value of the Ginzburg-Landau parameter $\kappa$ as a function of temperature, for the cases $q=4$, and $q=24$.}

\end{figure}

\begin{figure}[t!]
\begin{center}
\begin{picture}(250,150)
\put(10,0){\includegraphics*[width=0.5\linewidth,angle=0]{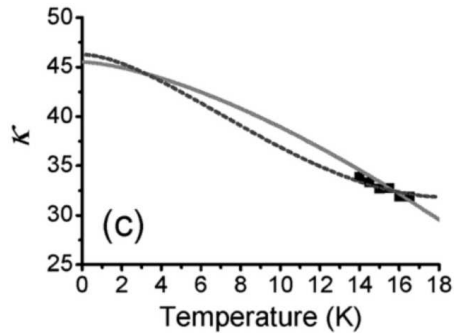}}%
\end{picture}
\caption{\label{korean} Temperature dependence of the Ginzburg-Landau parameter $\kappa$. The dashed line corresponds to the empirical curve of the form (\ref{fitk}) for the high-$T_{c}$ material $\text{Nb}_{3}\text{Sn}$. Figure taken from \cite{Kim:2007fj}. }
\end{center}
\end{figure}

Having determined the correlation length $\xi_{0}$, we can also calculate the remaining characteristic length of the superconductor, namely the Ginzburg-Landau penetration length $\lambda$. This can be done directly from its definition as (\ref{Lambda})
\begin{equation}
\lambda^{2}=\frac{1}{4\pi \,q^{2}n_{s}}\,,
\end{equation}
or, in terms of $\tilde{n}_{s}$
\begin{equation}
\lambda^{2}=\frac{1}{4\pi \tilde{n}_{s}}\,,
\end{equation}
where, as we have seen, $\tilde{n}_{s}$ is given holographically by (\ref{nstilde}). In figure (\ref{lambda4}) we show its behaviour as a function of temperature, for the $q=4$ case.  With both characteristic lengths, we can consequently obtain numerical values for the Ginzburg-Landau parameter, defined as $\kappa = \lambda/\xi$ (see (\ref{kapp})). We note that the definition of $\kappa$ uses the Ginzburg-Landau coherence length $\xi$, which is related to the superconducting coherence length calculated above by $\xi^{2}=2 \xi_{0}^{2}$. We obtain
\begin{equation}
\label{holok}
\kappa=\sqrt{\frac{1}{8\pi\,\tilde{n}_{s}\,\xi_{0}^{2}}}\,.
\end{equation}
The behaviour of $\kappa$ as a function of temperature is shown in figures (\ref{kg4}) and (\ref{kg24}) for the cases $q=4$ and $q=24$, respectively. A striking feature concerning the large-$q$ Ginzburg-Landau parameter, like the $q=24$ case presented in (\ref{kg24}), is that its qualitative behaviour can be modeled using the same kind of empirical fitting already used for high-$T_{c}$ superconducting material $\text{Nb}_{3}\text{Sn}$ in \cite{summers}, where the authors determined the temperature dependence for $\kappa$ to be given by
\begin{equation}
\label{fitk}
\kappa(T)=\kappa(0)\left(a_{0}-b_{0} (T/T_{c})^{2}\left(1-c_{0}\, \log(T/T_{c})\right)\right)\,,
\end{equation}
with $a_{0}$, $b_{0}$ and $c_{0}$ given empirically. This is shown in figure (\ref{korean}). This curve has the same shape of figure (\ref{kg24}). Indeed, the same formula can be used to fit our results to very good approximation, giving rise to the essentially same plot shown in figure (\ref{kg24}). The same can be done with the other large-$q$ cases.

\bigskip

\begin{figure}[t!]
\begin{center}
\begin{picture}(250,170)
\put(10,0){\includegraphics*[width=.6\linewidth,angle=0]{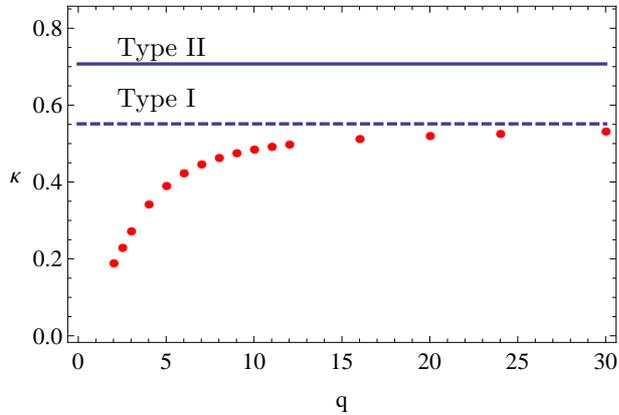}}%
\put(50,136){Type II}
\put(50,117){Type I}
\end{picture}
\caption{\label{kappaevolution} Evolution of the Ginzburg-Landau parameter $\kappa$ as a function of the scalar field charge $q$.}
\end{center}
\end{figure}

In figure (\ref{kappaevolution}) we show the evolution of $\kappa$ as the value of $q$ increases. The plot was made by taking the value of $\kappa$ closest to the critical temperature for each charge. We also show the line $\kappa=1/\sqrt{2}$ (bold line) corresponding to the value where, according to Ginzburg-Landau theory, the system turns from a type I to a type II superconductor. Since numerical factors have been maintained in our Ginzburg-Landau interpretation, this exact value still holds. What can be seen is that the system behaves as a type I superconductor, with the value of $\kappa$ increasing monotonically and approaching the asymptotic value $\kappa \approx 0.55 $, shown as a dashed line in figure (\ref{kappaevolution}), which is below $\kappa = 1/\sqrt{2}$.~~\footnote{
We note that the asymptotic constant behaviour of $\kappa$ as the value of $q$ grows can be seen directly from (\ref{holok}), where, making use of the fact that at the critical temperature $\xi_{0}=1/A_{q}\mathcal{O}_{3}$, we can write $\kappa$ as
\begin{equation}
\kappa=\sqrt{\frac{A_{q}^{2}\mathcal{O}_{3}^{2}}{8\pi\,\tilde{n}_{s}}}\,,
\end{equation}
and, using (\ref{numeq})
\begin{equation}
\kappa=\sqrt{\frac{C_{0}A_{q}^{2}\,T_{c}(q)}{8\pi\,q}}\,.
\end{equation}
Since for large $q$ we know that both $A_{q}$ and $T_{c}$ behave as $q^{1/3}$, then in that limit we will have $\kappa \sim \text{const.}$} 

\bigskip

\begin{figure}[t!]
\begin{center}
\begin{picture}(250,170)
\put(10,0){\includegraphics*[width=.6\linewidth,angle=0]{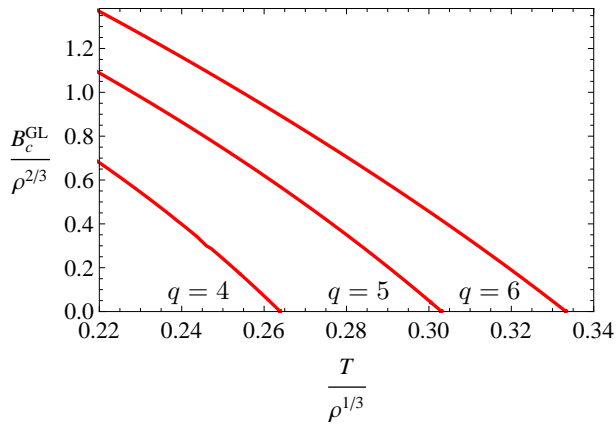}}%
\put(70,47){$q=4$}
\put(130,47){$q=5$}
\put(179,47){$q=6$}
\end{picture}
\caption{\label{magnet} Value of the Ginzburg-Landau critical magnetic field $B_{c}^{\text{GL}}$ as a function of temperature, for $q=4~,5~,6$.}
\end{center}
\end{figure}

\begin{figure}[t!]
\begin{center}
\begin{picture}(250,170)
\put(10,0){\includegraphics*[width=.6\linewidth,angle=0]{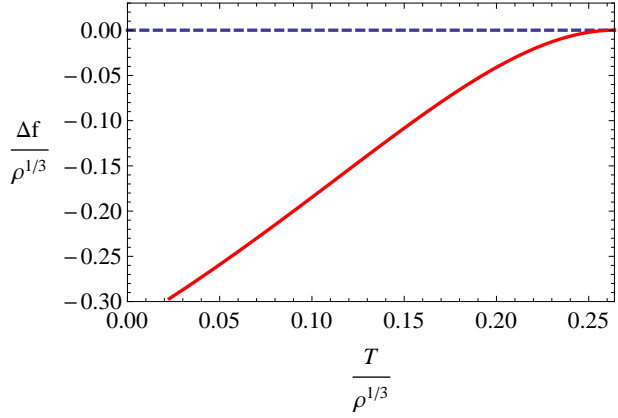}}%
\end{picture}
\caption{\label{DF4} Value of the Helmholtz free energy difference computed through standard holographic techniques $\Delta f$  as a function of temperature, for $q=4$.}
\end{center}
\end{figure}

\begin{figure}[t!]
\begin{center}
\begin{picture}(250,170)
\put(10,0){\includegraphics*[width=.6\linewidth,angle=0]{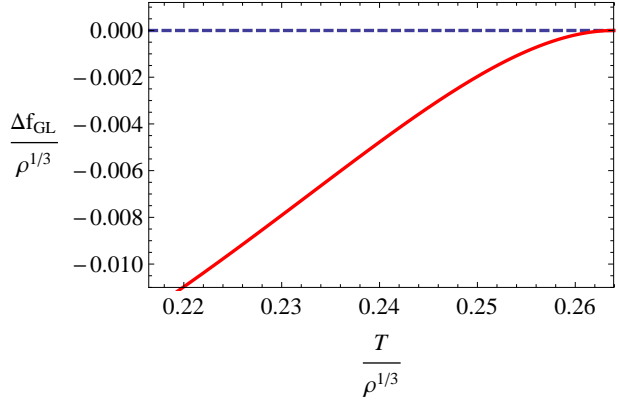}}%
\end{picture}
\caption{\label{DFGL4} Value of the Helmholtz free energy difference computed by the Ginzburg-Landau approach $\Delta f_{\text{GL}}$ as a function of temperature, for $q=4$.}
\end{center}
\end{figure}

An interesting fact about the phenomenological Ginzburg-Landau description is that, according to it, we can calculate the value of the critical magnetic field that breaks the superconducting phase of the theory. According to the Ginzburg-Landau theory, this critical field, which we will refer to as $B_{c}^{\text{GL}}$, is given by (\ref{HCritical})
\begin{equation}
\label{BGL1}
B_{c}^{\text{GL}}=\sqrt{4\pi}\frac{\left|\alpha\right|}{\sqrt{\beta}}\,,
\end{equation}
where we used the fact that for holographic superconductors $H=B/\mu_{0}$. It is important to notice that this critical field arises in Ginzburg-Landau theory from balancing the condensate part of the free energy against its purely magnetic part (see Appendix). This field points in the $x_{3}$-direction, and should be related to the real part of
\begin{equation}
F_{x_{1},x_{2}}=i\,k\,  A_{x}^{(0)}\,.
\end{equation}
After substitution of (\ref{alpha}) and (\ref{beta1}) in (\ref{BGL1}) we have
\begin{equation}
\label{BGL}
B_{c}^{\text{GL}}=\sqrt{\frac{\pi}{ q\, C_{0} T_{c}}}\frac{\mathcal{O}_{3}}{\xi_{0}}\,.
\end{equation}
In figure (\ref{magnet}) we show how this critical field behaves as a function of temperature for the cases $q=4,5,6$. Near $T_{c}$, using (\ref{xiTc}), the last expression becomes
\begin{equation}
\label{BGL2}
B_{c}^{\text{GL}}\approx\sqrt{\frac{\pi}{ q\, C_{0} T_{c}}}A_{q}\mathcal{O}_{3}^{2}\,,
\end{equation}
where we see that $B_{c}^{\text{GL}}$ has a near-$T_{c}$ behaviour $B_{c}^{\text{GL}}\sim \left(1-T/T_{c}\right)$, consistent with mean field theory. 

\bigskip

\begin{figure}[t!]
\begin{center}
\begin{picture}(250,170)
\put(10,0){\includegraphics*[width=.6\linewidth,angle=0]{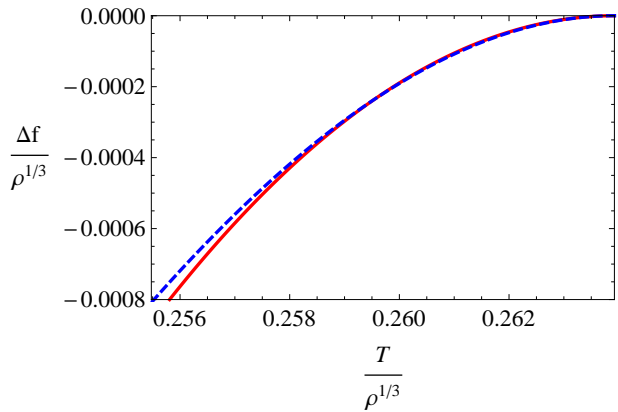}}%
\end{picture}
\caption{\label{FEcomp4} Comparison between free energy densities as a function of temperature, for $q=4$. The bold line corresponds to $\Delta f$, while the dashed line corresponds to $\Delta f_{\text{GL}}$.}
\end{center}
\end{figure}

\begin{figure}[t!]
\begin{center}
\begin{picture}(250,170)
\put(10,0){\includegraphics*[width=.6\linewidth,angle=0]{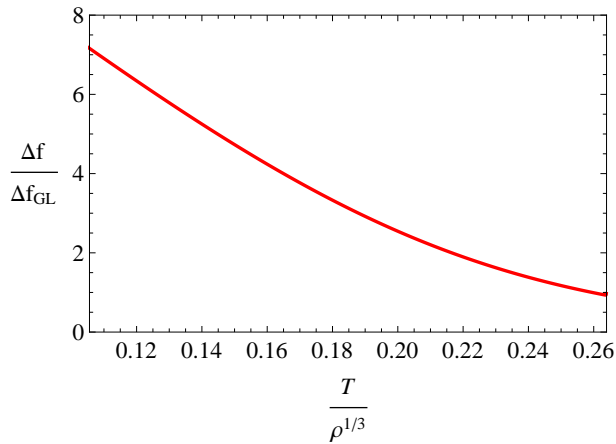}}%
\end{picture}
\caption{\label{DFratio4} Value of the ratio $\Delta f/\Delta f_{\text{GL}}$ as a function of temperature, for $q=4$.}
\end{center}
\end{figure}

\begin{figure}[t!]
\begin{center}
\begin{picture}(250,170)
\put(10,0){\includegraphics*[width=.6\linewidth,angle=0]{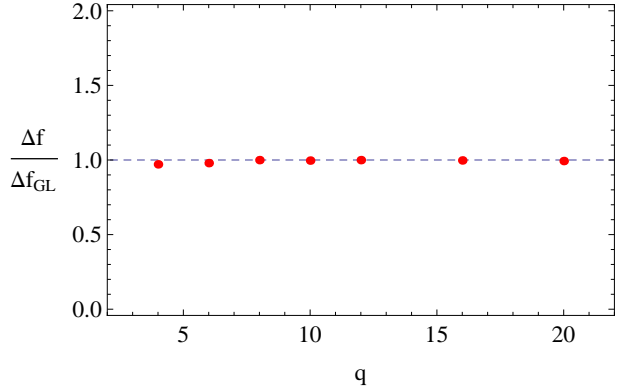}}%
\end{picture}
\caption{\label{FEratioq2} Value of the ratio $\Delta f/\Delta f_{\text{GL}}$ evaluated at $T=T_{c}$, for different values of $q$.}
\end{center}
\end{figure}

Finally, we want to see how our current Ginzburg-Landau approach holds up with regard to the Helmholtz free energy density of the system.\footnote{The Helmholtz free energy density is the appropriate thermodynamic potential in our case, given our choice to work with fixed charge density $\rho$, i.e. in the canonical ensemble.} The Helmholtz free energy density $f$ is given in general by
\begin{equation}
\label{FEdef}
f=\epsilon-Ts\,,
\end{equation}
where $\epsilon$ and $s$ are the total energy and entropy density, respectively. In order to calculate the Helmholtz free energy, we follow \cite{Hartnoll:2008kx} and make use of the fact that the stress-energy tensor must be traceless. For our particular case, this implies that $\epsilon=3P$, where $P$ is the pressure. Substituting in the thermodynamic identity $\epsilon = s T + \mu \rho -P$, and in the formal definition (\ref{FEdef}) we obtain the expression
\begin{equation}
f = \frac{1}{4}\left(3 \mu \rho - s T\right)\,,
\end{equation}
which is used to compute $f$ in both the condensed and normal phases, as a function of $T$ and for different values of $q$.  We focus on the free energy difference $\Delta f=f_{\text{sc}}-f_{\text{n}}$, where $f_{\text{sc}}$ corresponds to the free energy in the superconducting phase, while $f_{\text{n}}$ corresponds to the free energy in the normal phase. The free energy difference $\Delta f$ of the system is shown in figure (\ref{DF4}) as a function of temperature, for the particular case $q=4$.

Meanwhile, according to Ginzburg-Landau theory, the free energy difference is given by equation (\ref{LGFreeEnergy}) in the Appendix. Since we are working in the approximation where the order parameter $\left|\Psi\right|\approx\left|\Psi_{\infty}\right|$, near $T_c$ we can safely focus on the first two terms
\begin{equation}
\label{FEGL}
\Delta f_{\text{GL}}\approx\alpha \left|\Psi\right|^{2}+\frac{1}{2}\beta \left|\Psi\right|^{4}\,,\hspace{25pt} (T\approx T_{c})\,.
\end{equation}
Substituting in (\ref{FEGL}) the values obtained holographically earlier in this section for $\left|\Psi\right|$, $\alpha$ and $\beta$, we have
\begin{equation}
\label{DFGLeq}
\Delta f_{\text{GL}}=-\frac{1}{8\,q\, C_{0}T_{c}}\frac{\mathcal{O}_{3}^{2}}{\xi_{0}^{2}}\,.
\end{equation}
In figure (\ref{DFGL4}) we show the behaviour of $\Delta f_{\text{GL}}$ as a function of temperature, for the $q=4$ case. We then compare both free energy differences $\Delta f$ and $\Delta f_{\text{GL}}$. Figure (\ref{FEcomp4}) compares the free energies computed by the two different methods. We see that there is an excellent agreement, showing that both descriptions should be more accurate near the critical temperature. In figure (\ref{DFratio4}) we show the ratio $\Delta f/\Delta f_{\text{GL}}$ as a function of temperature, for the $q=4$. We find that the ratio reaches the constant value $\sim 0.99$ at $T=T_{c}$. Moreover, this value of the ratio at $T=T_{c}$ is found to be the same for all values of $q$ considered. This is shown in figure (\ref{FEratioq2}), where the value of the ratio $\Delta f/\Delta f_{\text{GL}}$ evaluated at $T_c$ is shown for different values of $q$.

\begin{figure}
\centering
\begin{subfigure}{.5\textwidth}
  \centering
 \begin{picture}(250,150)
\put(0,0){\includegraphics*[width=1\linewidth]{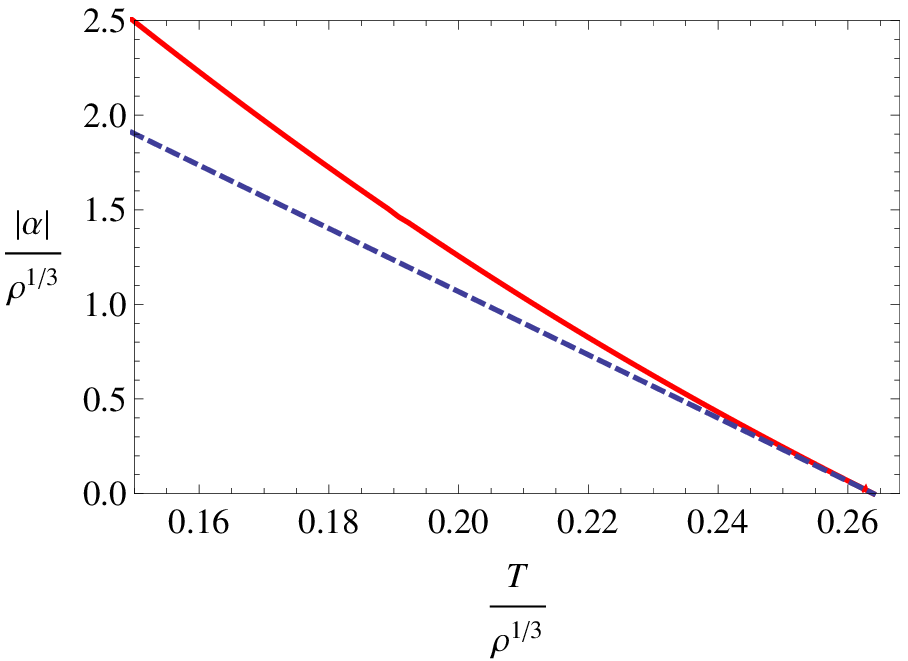}}%
\end{picture}
  \caption{$\alpha$}
  \label{alphaK}
\end{subfigure}%
\begin{subfigure}{.5\textwidth}
  \centering
	\begin{picture}(250,150)
\put(10,0){\includegraphics*[width=1\linewidth]{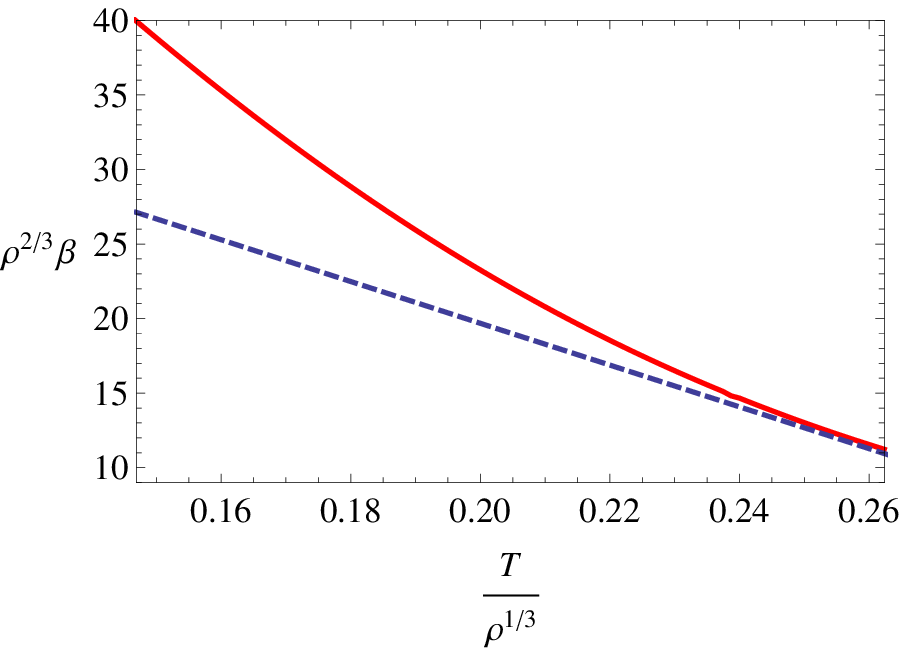}}%
\end{picture}
  \caption{$\beta$}
  \label{betaK}
\end{subfigure}
\caption{Near-$T_{c}$ comparison of the Ginzburg-Landau parameters $\alpha$ and $\beta$ obtained through the method developed in \cite{Herzog:2008he} (dashed line) and through our current Ginzburg-Landau approach (bold line), for the case $q=4$. The parameters computed through \cite{Herzog:2008he} are presented only to linear level in $T$.}
\end{figure}

\bigskip

It is interesting to compare the present results with the results of \cite{Herzog:2008he} for a $d=3+1$-bulk system. The authors in this paper, using a rather different method, performed a fit of the free energy using the Ginzburg-Landau form (\ref{FEGL}) with the corresponding order parameter $\mathcal{O}_i$. By doing this, they obtain near-$T_c$ expressions for $\alpha$ and $\beta$ as functions of temperature which agree with the results of standard Ginzburg-Landau theory. Applying the same procedure to fit the free energy in our $d=4+1$ bulk dimensional system\footnote{I order to apply the methods developed in \cite{Herzog:2008he}, we note that our system is a $d=4+1$ dimensional version of the model they work with, with no spatial component of the gauge field (superfluid velocity $\xi=0$, in the authors notation), and that we are working in the canonical ensemble while in \cite{Herzog:2008he} the authors consider the grand canonical ensemble.}, we find that, in the particular $q=4$ case, the Ginzburg-Landau parameters $\alpha$ and $\beta$ behave near-$T_{c}$ and up to lineal level in $T$ as
\begin{equation}
\label{kovtun}
\left|\alpha\right| = 4.41 \left(1-T/T_{c}\right)\,,\hspace{20pt}\beta = 10.95 + 36.75\left(1-T/T_{c}\right)\,,
\end{equation}
Meanwhile, in our current Ginzburg-Landau approach, the parameters $\alpha$ and $\beta$, which are computed through equations (\ref{alpha}) and (\ref{beta1}) respectively, can be expressed near-$T_{c}$ and up to linear level in $T$ as
\begin{equation}
\label{GLlinear}
\left|\alpha_{\text{GL}}\right| = 4.45
 \left(1-T/T_{c}\right)\,,\hspace{20pt}\beta_{\text{GL}} = 11.23 + 35.2\left(1-T/T_{c}\right)\,.
\end{equation}
Comparing (\ref{kovtun}) and (\ref{GLlinear}) we see that near $T_{c}$ both results are quantitatively very similar. In figures (\ref{alphaK}) and (\ref{betaK}) we show how the expressions (\ref{kovtun}) for $\alpha$ and $\beta$ obtained through the methods used in \cite{Herzog:2008he} compare near-$T_{c}$ with the parameters computed by our Ginzburg-Landau approach. Observing this good agreement between both results, we conclude that the methods developed in \cite{Herzog:2008he} and in this paper can be viewed as complementary. We notice that, in the Ginzburg-Landau approach, the whole functional dependency of $\alpha$, $\beta$ and the free energy on $T$ is contained entirely on simple combinations of $\mathcal{O}_{3}^{2}$ and $\xi_{0}^{2}$, which arise naturally when looking for consistency.


\bigskip

\pagebreak

\section{Constant external magnetic field}

\subsection{A constant magnetic field background}

We will now introduce a uniform external magnetic field into our model. To do this, we use the procedure described in \cite{D'Hoker:2009bc} to build perturbatively an asymptotically-$AdS$ fixed magnetic background. The starting point is a $d=4+1$ Einstein-Maxwell action with a negative cosmological constant 
\begin{equation}
S=\int d^{5}x \sqrt{-g}\left(R+\frac{12}{L^{2}}-\frac{1}{4} F^{2}\right).
\end{equation}
We consider a magnetic ansatz for the gauge field
\begin{equation}
\label{gansatz}
A = \phi(r) dt + \frac{B}{2}\; \left(-x_{2} d x_{1} + x_{1} dx_{2}\right)\,,
\end{equation}
which means that we will have a constant external magnetic field pointing in the $x_{3}$-direction of the dual field theory, given by $F_{x_{1},x_{2}}=B$. For the metric, we propose the ansatz
\begin{equation}
 \label{metricansatz}
ds^{2} = - g(r) dt^{2} + \frac{dr^{2}}{g(r)}+ e^{2 V(r)} \left(dx_{1}^{2} + dx_{2}^{2}\right) + e^{2 W(r)}dx_{3}^{2}\,.
\end{equation}
Such an ansatz has a $SO(2)$ isometry in the $x_{1}-x_{2}$ plane, and is invariant under translations in the $x_{3}$ direction, due to the fact that the magnetic field will define a preferred direction in the $\left(x_{1},~x_{2},~x_{3}\right)$ space. We will look for asymptotically $AdS$ black hole solutions for the metric. The Einstein equations for this system are
\begin{equation}
R_{\mu\nu}+g_{\mu\nu}\left(\frac{1}{12}F^{2}+\frac{4}{L^{2}}\right)+\frac{1}{2}F_{\mu}^{\;\;\lambda}F_{\nu\lambda}=0\,.
\end{equation}
Substituting the ansatz (\ref{gansatz}) and (\ref{metricansatz}) into these equations, we get
\begin{eqnarray}
2 V'^{2}+W'^{2}+2 V'' + W''&=&0\,,\\
\frac{B^{2}}{2}e^{-4 V}+\left(g\left(V-W\right)'\right)'+g\left(2 V -W\right)'\left(V-W\right)'&=&0\,,\\
-\frac{B^{2}}{3}e^{-4 V}-\frac{2}{3}\phi'^{2}-\frac{8}{L^{2}}+g' \left(2 V + W\right)'+g''&=&0\,,
\end{eqnarray}
while the gauge field equation is given by
\begin{equation}
\left(2 V + W\right)'\phi'+\phi'' =0\,.
\end{equation}

One then considers the following expansion in powers of $B$ around $B=0$, up to second order:
\begin{eqnarray}
g(r) &=& g_{0}(r) + B^{2} g_{2}(r)+\ldots\\
V(r) &=& V_{0}(r) + B^{2} V_{2}(r)+\ldots\\
W(r) &=& W_{0}(r) + B^{2} W_{2}(r)+\ldots\\
\phi(r)&=&\phi_{0}(r) + B^{2} \phi_{2}(r)+\ldots\,.
\end{eqnarray}
As described in \cite{D'Hoker:2009bc}, this expansion is reliable for $B \ll T^{2}$. The $B^{0}$-order equations are solved  by the usual $AdS$ Reissner-Nordstr\"{o}m solution:
\begin{eqnarray}
\phi_{0}(r)&=& \frac{1}{2} - \frac{\rho}{r^{2}}\,,\\
g_{0}(r) &=& \frac{r^{2}}{L^{2}} + \frac{\rho^{2}}{3 r^{4}} - \frac{3 r_{h}^{6}+L^{2}\rho^{2}}{3 L^{2} r_{h}^{2} r^{2}}\,,\\
V_{0}(r) &=& W_{0}(r) = \log r\,.
\end{eqnarray}
From now on, we will set $L=1$, following our previous convention. The $B^2$-order equations are:
\begin{eqnarray}
\left(r^2 \left(2 V_{2} + W_{2}\right)'\right)'=0\,,\label{T2}\\
\frac{1}{2r}+\left(r^3 g_{0} \left(V_{2}-W_{2}\right)'\right)'=0\,,\label{S2}\\
-\frac{1}{3r}+\left(r^{3}g'_{2}\right)' + r^3 g'_{0} \left(2 V_{2}+ W_{2}\right)'=0\,,\label{g2}\\
2\rho \left(2 V_{2}+W_{2}\right)'+\left(r^{3} \phi'_{2}\right)'=0\,.\label{phi2}
\end{eqnarray}

\bigskip

From (\ref{T2}), demanding that $V_{2}$ and $W_{2}$ vanish at infinity and be regular at the horizon, we obtain 
\begin{equation}
\label{V2W2}
2V_{2}+W_{2}=0.
\end{equation}
Substituting this result in (\ref{phi2}), and demanding that $\phi_{2}$ vanishes at both the horizon and infinity, we have $\phi_{2}=0$. Also, from (\ref{g2}) and demanding that $g_{2}$ vanishes also at the horizon and infinity, the solution for $g_{2}$ is
\begin{equation}
g_{2}(r)=-\frac{1}{6 r^{2}}\log\left(\frac{r}{r_{h}}\right)\,.
\end{equation}
Finally, from equation (\ref{S2}) we get
\begin{equation}
V_{2}(r)=-\frac{1}{6}\int_{\infty}^{r} dr' \frac{\log\left(r'/r_{h}\right)}{r'^{3}g_{0}(r')}\,.
\end{equation}
and $W_{2}$ given by (\ref{V2W2}). From the solution up to second order in $B$ for $g(r)$ 
\begin{equation}
g(r)=r^{2}+\frac{\rho^{2}}{3\,r^{4}}-\frac{3\,r_{h}^{6}+\rho^{2}}{3r^{2}r_{h}^{2}}-\frac{B^{2}\log\left(r/r_{h}\right)}{6\,r^{2}}\,,
\end{equation}
we can obtain the Hawking temperature of the system
\begin{equation}
\label{ThawkingB}
T_{H}=\frac{24\,r_{h}^{6}-4\rho^{2}-B^{2}\,r_{h}^{2}}{24 \pi \,r_{h}^{5}}\,.
\end{equation}
Since we will continue to work in the canonical ensemble, we will set $\rho=1$ for the remainder of this section.

\subsection{Droplet solution and critical magnetic field}

We will now turn on a small scalar field in the fixed background given by the solutions constructed in the previous subsection. This will be analogous to the analysis made by \cite{Hartnoll:2008kx, Albash:2008eh} in a $d=3+1$ $AdS$. (For other, less conventional models, see e.g. \cite{Kuang:2013oqa}.) We propose an ansatz for the scalar field
\begin{equation}
\Psi(r,u)=\frac{1}{\sqrt{2}} R(r)\,U(u)\,,
\end{equation}
where we have made the change to cylindrical coordinates $dx_{1}^{2}+dx_{2}^{2}=du^{2}+u^{2}d\theta^{2}$. The equation (\ref{generalpsieq}) turns to be separable in this case, resulting in the equations
\begin{eqnarray}
U''+\frac{1}{u}U'+\left(\lambda-B^{2}q^{2}u^{2}\right)U=0\,,\label{Ueq}\\
R''+\left(\frac{g'}{g}+\frac{3}{r}\right)R'+\frac{1}{g}\left(\frac{q^{2}\phi^{2}}{g}-e^{-2\,V}\lambda-M^{2}\right)R=0\,,\label{Req}
\end{eqnarray}
where $\lambda$ is the separation constant, and must be equal to $\lambda_{n}=n\,q\,B$ in order for $U(u)$ to be finite as $u\rightarrow\infty$. We choose the $n=1$ mode, since this corresponds to the most stable solution \cite{Hartnoll:2008kx, Albash:2008eh}. In this case, the solution for (\ref{Ueq}) is a gaussian function
\begin{equation}
U(u)=\exp\left(-\frac{q\,B}{4}\,u^{2}\right)\,,
\end{equation}
which is the same result obtained in \cite{Hartnoll:2008kx} for a $d=3+1$-dimensional bulk. 

\begin{figure}
\centering
\begin{subfigure}{.5\textwidth}
  \centering
 \begin{picture}(250,150)
\put(0,0){\includegraphics*[width=1\linewidth]{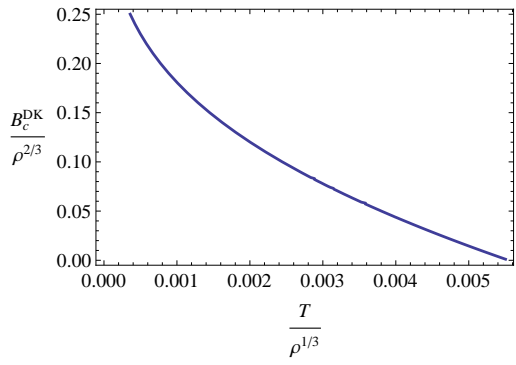}}%
\end{picture}
  \caption{ $q=1$}
  \label{B1}
\end{subfigure}%
\begin{subfigure}{.5\textwidth}
  \centering
	\begin{picture}(250,150)
\put(10,0){\includegraphics*[width=1\linewidth]{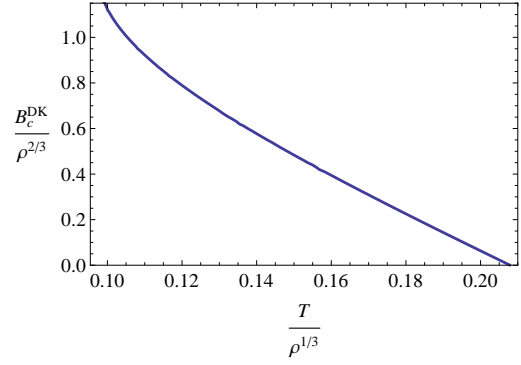}}%
\end{picture}
  \caption{ $q=3$}
  \label{B3}
\end{subfigure}
\begin{subfigure}{.5\textwidth}
  \centering
 \begin{picture}(250,150)
\put(0,0){\includegraphics*[width=1\linewidth]{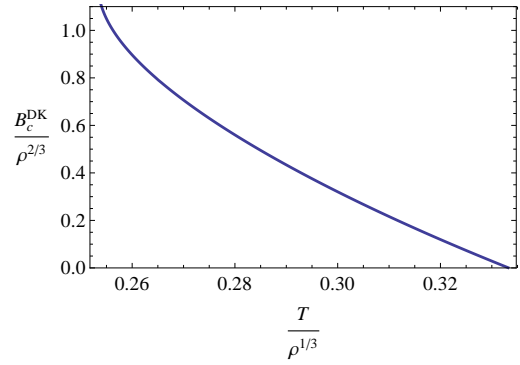}}%
\end{picture}
  \caption{ $q=6$}
  \label{B6}
\end{subfigure}%
\caption{Value of the critical magnetic field $B_{c}^{\text{DK}}$ as a function of temperature, for different values of $q$.}
\end{figure}

Substituting $\lambda_{1}$ in (\ref{Req}) and changing to the $z=r_{h}/r$ coordinate, we get
\begin{equation}
\label{Reqz}
R''+\left(\frac{g'}{g}-\frac{1}{z}\right)R'+\frac{r_{h}^{2}}{g\,z^{4}}\left(\frac{q^{2}\phi^{2}}{g}-2\,q\,B\,e^{-2\,V}-M^{2}\right)R=0\,,
\end{equation}
from where we derive the boundary regularity condition
\begin{equation}
R'(1)=\frac{r_{h}^{2}}{g'(1)}\left(2\,q\,B\,e^{-2 V(1)}+M^{2}\right)\,R_{0}\,,
\end{equation}
where $R_{0}=R(1)$. Again, we choose $M^{2}=-3$, which gives the asymptotic behaviour
\begin{equation}
R=\mathcal{O}_{1}\frac{z}{r_{h}}+\mathcal{O}_{3}\frac{z^{3}}{r_{h}^{3}}+\ldots\,.
\end{equation}
Since we will not be concerned about the absolute normalization of $\mathcal{O}_{3}$, we will take advantage of the linearity of (\ref{Reqz}) and set $R_{0}=1$. This will leave $B$ and $r_{h}$ as the only input parameters in the equation. As in the previous section, we will choose to set $\mathcal{O}_{1}=0$ and solve the differential equation (\ref{Reqz}) enforcing this choice through the shooting method. This leaves $r_{h}$, and therefore $T_{H}$ in (\ref{ThawkingB}),  as the only free parameter of the system and will allow us to determine the value of $B$ as a function of temperature. This magnetic field $B_{c}$ will correspond to the value above which superconductivity is broken. From the holographic point of view, the critical magnetic field obtained above measures an instability of the bulk scalar field $\psi$. Indeed, from the effective mass of the scalar field
\begin{equation} 
M_{\text{eff}}^{2}=M^{2}-\frac{q^{2}}{g}\Phi^{2}+\frac{q^{2}}{4}e^{-2V}u^{2}B^{2}\,,
\end{equation}
we see that the magnetic term has an opposite sign to the electric term, which is responsible for lowering the effective mass below the Breitenlohner-Freedman bound and making the field tachyonic. The sign difference means then that the magnetic term lowers the critical temperature under which the scalar field becomes unstable \cite{Albash:2009iq}. We will refer to the critical magnetic field obtained in this section as $B_{c}^{\text{DK}}$, in order to distinguish it from the critical magnetic field as given by Ginzburg-Landau theory, $B_{c}^{\text{GL}}$, which was introduced in the preceding section.

\bigskip

In figures (\ref{B1})-(\ref{B6}) we show the value of the critical magnetic field $B_{c}^{\text{DK}}$ for the cases $q=1\,,3\,,6$. We only show the region near the critical temperature where our approximation is valid. The divergence of $B_{c}^{\text{DK}}$ as the temperature moves away from $T_{c}$ is typical of the no-backreaction approach we are using, as observed in \cite{Albash:2009iq}.

\begin{figure}[t!]
\begin{center}
\begin{picture}(250,160)
\put(0,0){\includegraphics*[width=.6\linewidth,angle=0]{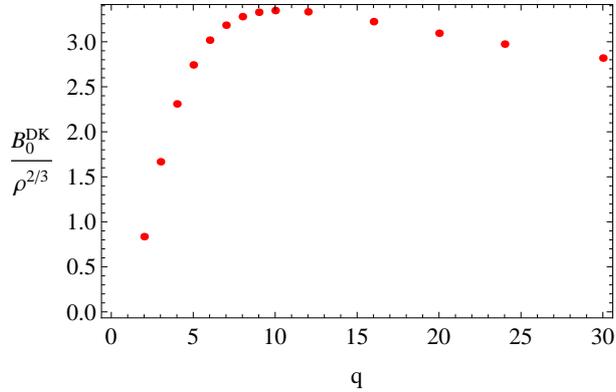}}%
\end{picture}
\caption{\label{BDK0fig} Behaviour of the near-$T_{c}$ coefficient $B_{0}^{\text{DK}}$ as a function of the scalar field charge $q$.}
\end{center}
\end{figure}

Finally, we find numerically that near-$T_{c}$ the critical magnetic field $B_{c}^{\text{DK}}$ behaves as
\begin{equation}
\label{BDK0}
B_{c}^{\text{DK}}\sim B_{0}^{\text{DK}}\left(1-T/T_{c}\right)\,,
\end{equation}
in accordance to mean field theory. The behaviour of the factor $B_{0}^{\text{DK}}$ as a function of the scalar field charge $q$ is shown in figure (\ref{BDK0fig}). For large $q$, one finds numerically that $B_{0}^{\text{DK}}\sim q^{-1/3}$. 

\bigskip

\begin{figure}[t!]
\begin{center}
\begin{picture}(250,160)
\put(0,0){\includegraphics*[width=.6\linewidth,angle=0]{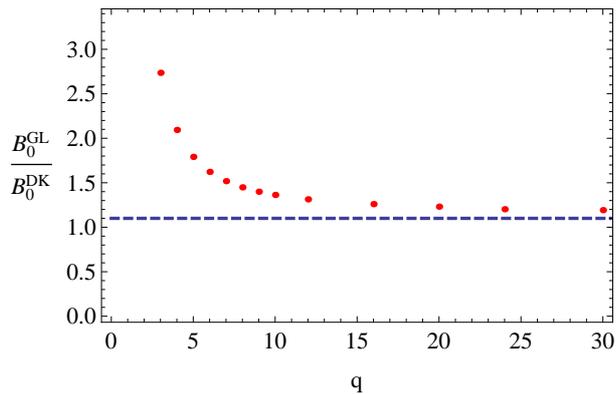}}%
\end{picture}
\caption{\label{magrat} Behaviour of the ratio $B_{0}^{\text{GL}}/B_{0}^{\text{DK}}$ as a function of the scalar field charge $q$. The dashed line corresponds to the asymptotic limit $\sim 1.1$}
\end{center}
\end{figure}

It is interesting to note that the critical magnetic fields $B_{c}^{\text{GL}}$ and $B_{c}^{\text{DK}}$ measure different aspects of the response of the system to a magnetic field: with $B_{c}^{\text{DK}}$ measuring an instability in the scalar bulk field, and $B_{c}^{\text{GL}}$ arising from a balancing between the condensate part and the purely magnetic part of the free energy according to Ginzburg-Landau theory.\footnote{I wish to thank the referee for pointing out this distinction.} We found that near $T_{c}$ both critical magnetic fields behave as $\sim \left(1-T/T_{c}\right)$. Explicitly
\begin{equation}
B_{c}^{\text{DK}}\sim B_{0}^{\text{DK}}\left(1-T/T_{c}\right)\,,\hspace{25pt} B_{c}^{\text{GL}}\sim B_{0}^{\text{GL}}\left(1-T/T_{c}\right)\,,
\end{equation}
with
\begin{equation}
B_{0}^{\text{GL}}\equiv \sqrt{\frac{\pi}{ q C_{0}T_{c}}}A_{q}\mathcal{O}_{0}^{2}\,.
\end{equation}
(See equation (\ref{BGL2}).) Since we know that for large $q$ we have $A_{q}\sim q^{1/3}$, $T_{c}\sim q^{1/3}$ and $\mathcal{O}_{0}\sim q^{0}$, then we conclude that, in this limit, $B_{0}^{\text{GL}}\sim 1/q^{1/3}$ (or equivalently, $B_{0}^{\text{GL}}\sim 1/T_{c}$) and thus, we find that both $B_{0}^{\text{DK}}$ and $B_{0}^{\text{GL}}$ have the same large-$q$ behaviour. Indeed, this can be seen in figure (\ref{magrat}), where we show the ratio $B_{0}^{\text{GL}}/B_{0}^{\text{DK}}$ as a function of the scalar field charge $q$, and where we find numerically that it tends asymptotically to the constant value $\sim 1.1$.

\section{Conclusions}

In this paper we have constructed a family of minimal holographic superconducting models in $d=4+1$ $AdS$ spacetime, characterized by their scalar field charge $q$ (or, equivalently, by their critical temperature $T_{c}$). We have first turned on a small magnetic perturbation in the $x_{1}$ component of the gauge field, as well as a small perturbation of the scalar field around the condensed solution. By making a Ginzburg-Landau phenomenological interpretation of the dual field theory, we calculated the Ginzburg-Landau parameters and characteristic lengths as a function of temperature. We found that they have a behaviour consistent with that of usual superconducting systems as described by mean field theory. We also calculated the Ginzburg-Landau parameter $\kappa$ for different values of the scalar field charge $q$. From this calculation we find that, as the value of $q$ increases, the Ginzburg-Landau parameters approaches asymptotically the value $\kappa \sim 0.55 <1/\sqrt{2}$. From this we can conclude that the system will behave as a type I superconductor for all values of $q$ considered. We have also calculated the Helmholtz free energy density of the system using the proposed Ginzburg-Landau approach, and compared it with the free energy computed with the standard holographic techniques. It was found that both approaches are consistent near $T_{c}$. Also, through calculations of the free energy of the system, the Ginzburg-Landau approach was compared with the method developed in \cite{Herzog:2008he} for calculating the parameters $\alpha$ and $\beta$. Both methods were shown to be in excellent agreement.

\bigskip

Next, we turned off the magnetic fluctuation and probed our system with a constant magnetic field $B$. This was done by using the black brane solution of \cite{D'Hoker:2009bc} in $d=4+1$ $AdS$ up to order $B^{2}$. With this perturbative solution, we showed the formation of droplet condensate solutions in this fixed background and calculated the critical magnetic field above which the superconducting phase is broken. The field obtained in this fashion was compared with the critical magnetic field obtained in the Ginzburg-Landau approach. While both fields measure different aspects of the response of the system to a magnetic field, we found that near $T_c$ both fields behave as $B_{c}\sim B_{0}\left(1-T/T_c\right)$ and that their corresponding factors $B_{0}$ behave as $\sim 1/q^{1/3}$ (or equivalently as $\sim 1/T_c$) for large $q$.

\bigskip

One of the main results of this paper is to show that a very simple phenomenological model in $d=4+1$ $AdS$ spacetime allows for a consistent Ginzburg-Landau description of the boundary theory, where all the Ginzburg-Landau parameters and characteristic lengths can be calculated using holographic methods, and whose behaviour is in accordance to the one predicted by traditional mean field theory. Moreover, we also observe that, as the value of the scalar field charge $q$ increases, the Ginzburg-Landau parameter of the model tends asymptotically to a well defined value that characterizes the dual superconducting system as type I. In this respect it is natural to ask how the Ginzburg-Landau parameter obtained in this paper could change by modifying the model by, for instance, changing the value of the bulk-scalar field mass $M^{2}$, the quantization condition at the boundary, or using higher order corrections in $\psi$ for the potential as in top-down approaches. All these questions call for further research.

\bigskip

\textbf{Note added}: As mentioned in the introduction, after the completion of this work, a paper \cite{Dias:2013bwa} appeared where the Ginzburg-Landau parameter is computed by a different method for $2+1$ dimensional superconductors. The authors find that the system is type II at lower charge and type I at higher charge. In the $3+1$ dimensional holographic superconductor studied here, we find that the system behaves as type I for all values of the charge.

\section*{Acknowledgements}

It is a pleasure to thank J. G. Russo for his guidance and encouragement during the research of this project, and F. Aprile for his valuable comments and suggestions. I also wish to thank the referee for important clarifications and corrections. This research was funded by CONACyT grant No.306769.

\pagebreak

\begin{appendices}

\section{Review of Ginzburg-Landau theory }

In this section we will review very briefly the main aspects of the Ginzburg-Landau model of superconductivity. (See, for example \cite{tinkham}.) The Helmholtz free energy density difference of the theory is given by
\begin{equation}
\label{LGFreeEnergy}
 \Delta f = \alpha(T) \left|\Psi\right|^{2}+\frac{1}{2}\beta(T) \left|\Psi\right|^{4}+\frac{\hbar^{2}}{2 m}\left|\left(\nabla-\frac{ i q}{\hbar } \textbf{A}\right)\Psi\right|^{2}+\frac{B^{2}}{2\mu_{0}}\,,
\end{equation}
where $\Delta f=f_{\text{sc}}-f_{\text{n}}$, with $f_{\text{sc}}$ and $f_{\text{n}}$ being the free energy densities in the superconducting and normal phases of the system, respectively. Also, $\left|\Psi\right|^{2}$ is the order parameter of the theory and $\alpha$, $\beta$ and $\gamma$ are phenomenological parameters that have a temperature dependence in general. We have added a gauge field $A_{i}$ and the corresponding magnetic energy in order to describe a charged system. As explained in the paper, the superconducting carrier charge is consistently identified with the bulk scalar field charge $q$. We will adopt the usual convention $\alpha<0$, $\beta>0$.

When the external field and gradients are negligible, the free energy density difference (\ref{LGFreeEnergy}) can be approximated by
\begin{equation}
 \Delta f = \alpha \left|\Psi\right|^{2}+\frac{1}{2}\beta \left|\Psi\right|^{4}\,,
\end{equation}
which is minimized at
\begin{equation}
\label{PsiMinimum}
\left|\Psi_{\infty}\right|=\sqrt{\frac{\left|\alpha\right|}{\beta}}\,.
\end{equation}
Since deep inside the superconductor the external fields and gradients can be neglected, the critical parameter $\Psi$ will approach the value $\Psi_{\infty}$ as it goes deeper into the volume of the system. Inserting this value back in (\ref{LGFreeEnergy}), we get inside the material
\begin{equation}
\label{DFpsi}
\Delta f = -\frac{\alpha^{2}}{2 \beta}\,. 
\end{equation}
This last equation can be related to the critical magnetic field $H_{c}$, which is the value of the magnetic field needed to be applied to the system in a condensed phase in order to break superconductivity. Indeed, this field is determined by the specific magnetic energy density that needs to be added to the condensation energy to take the system into the normal phase, that is
\begin{equation}
f_{\text{sc}}+\frac{\mu_{0}}{2}H_{c}^2=f_{\text{n}}\,,
\end{equation}
or, equivalently
\begin{equation}
\label{DFHc}
\Delta f = -\frac{\mu_{0}}{2}H_{c}^2\,.
\end{equation}
Equating (\ref{DFpsi}) and (\ref{DFHc}), we obtain
\begin{equation}
\label{HCritical}
H_{c}^{2}=\frac{\alpha^{2}}{\mu_{0}\beta}\,.
\end{equation}
For values of $H>H_c$ it will be energetically more favorable for the system to be in the normal phase.

\bigskip

Going back to (\ref{LGFreeEnergy}), minimizing with respect to $\mathbf{A}$ and using $\nabla \times \mathbf{B} = \mu_{0}\mathbf{J}$ we arrive at
\begin{equation}
\label{LondonCurrent}
\mathbf{J}=-\frac{q^{2}}{m} \left|\Psi\right|^{2}\mathbf{A}\,.
\end{equation}
This is the well known \textit{London current}, which can also be derived from the phenomenological London theory, which gives
\begin{equation}
\label{londoneqns}
\mathbf{J}=-\frac{q^{2}}{m}\,n_{s} \mathbf{A}\,,
\end{equation}
where $n_{s}$ is the number density of superconducting electrons. Comparing this expression with (\ref{LondonCurrent}) we get a relation between $\left|\Psi\right|$ and $n_{s}$
\begin{equation}
\label{psins}
 \left|\Psi\right|^{2}= n_{s}\,.
\end{equation}

Finally, we can arrive at the following equation
\begin{equation}
\label{BEquation}
\nabla^{2}\mathbf{B}=\frac{1}{\lambda^{2}}\mathbf{B}\,,
\end{equation}
which has magnetic field solutions that decay exponentially inside the superconductor, with decay length $\lambda$, called the \textit{penetration length}, and given by
\begin{equation}
\label{Lambda}
\lambda^{2}=\frac{m}{\mu_{0}q^{2} n_{s}}.
\end{equation}
This length corresponds to the inverse mass of the gauge field after symmetry breaking. Combining (\ref{PsiMinimum}), (\ref{HCritical}) and (\ref{Lambda}), we arrive at the following expressions for $\alpha$ and $\beta$
\begin{eqnarray}
\alpha&=&-\frac{q^{2}\mu_{0}^{2}}{m} H_{c}^{2} \lambda^{2}\,,\\
\beta&=&\frac{q^{4}\mu_{0}^{3}}{m^{2}} H_{c}^{2} \lambda^{4}\,.
\end{eqnarray}

\bigskip

Minimizing (\ref{LGFreeEnergy}) with respect to $\Psi^{*}$, one has
\begin{equation}
\alpha \Psi + \beta \left|\Psi\right|^{2}\Psi-\frac{\hbar^{2}}{2 m} \left(\nabla - \frac{ i q}{\hbar } \mathbf{A}\right)^{2}\Psi=0\,.
\end{equation}
Then, when $\mathbf{A}=0$ we have
\begin{equation}
\label{LGPsiEquationI}
\alpha \Psi + \beta \Psi^{3} -\frac{\hbar^{2}}{2 m} \Psi''=0\,,
\end{equation}
where for simplicity we assumed that $\Psi$ is real and only depends on the dimension $x$. Expanding around the minimum as
\begin{equation}
\Psi(x)=\sqrt{\frac{\left|\alpha\right|}{\beta}} + \eta(x)\,,\;\;\;\;\;\; \left|\eta\right|\ll 1\,,
\end{equation}
and inserting in (\ref{LGPsiEquationI}), we have, up to second order the equation
\begin{equation}
\label{LGPsiEquationII}
2\left|\alpha\right|\eta-\frac{\hbar^{2}}{2 m} \eta''=0\,,
\end{equation}

which has the physical solution
\begin{equation}
\eta(x)\sim e^{-\frac{\left|x\right|}{\xi_{0}}}\,,
\end{equation}
where $\xi_{0}$, defined as
\begin{equation}
\label{correl}
\xi_{0}^{2}=\frac{\hbar^{2}}{4 m \left|\alpha\right|}\,,
\end{equation}
is the superconductor \textit{correlation length}, and it is a measure of the spatial decay of a small perturbance of $\Psi$ from its equilibrium value. It is customary, however, to work with the \textit{Ginzburg-Landau correlation length} $\xi$, given by $\xi^{2}=2\,\xi_{0}^{2}$, that is
\begin{equation}
\label{GLcorrel}
\xi^{2}=\frac{\hbar^{2}}{2 m \left|\alpha\right|}\,.
\end{equation}

\bigskip

Finally, from the characteristic lengths $\lambda$ and $\xi$ one can construct the \textit{Ginzburg-Landau parameter}, defined as:
\begin{equation}
\label{kapp}
\kappa = \frac{\lambda}{\xi}\,,
\end{equation}
whose value, based on surface energy calculations (see \cite{tinkham}), characterizes the behaviour of the system in a superconducting phase as:
\begin{eqnarray}
\kappa < \frac{1}{\sqrt{2}}\;\;\;\;\;\;\;\;\;\; \text{Type I Superconductor}\\
\kappa > \frac{1}{\sqrt{2}}\;\;\;\;\;\;\;\;\;\; \text{Type II Superconductor}
\end{eqnarray}
where a type II superconductor is one which allows partial penetration of a magnetic field, while a type I superconductor is one where the magnetic field is fully expelled from its volume by the Meissner effect.

\end{appendices}



\begin{thebibliography}{9}

\bibitem{Aharony:1999ti}
  O.~Aharony, S.~S.~Gubser, J.~M.~Maldacena, H.~Ooguri and Y.~Oz,
  Phys.\ Rept.\  {\bf 323} (2000) 183
  [hep-th/9905111].


\bibitem{Hartnoll:2008kx} 
  S.~A.~Hartnoll, C.~P.~Herzog and G.~T.~Horowitz,
  JHEP {\bf 0812}, 015 (2008)
  [arXiv:0810.1563 [hep-th]].
	
\bibitem{Hartnoll:2009sz}
  S.~A.~Hartnoll,
  Class.\ Quant.\ Grav.\  {\bf 26} (2009) 224002
  [arXiv:0903.3246 [hep-th]].
	
\bibitem{Horowitz:2010gk}
  G.~T.~Horowitz,
  Lect.\ Notes Phys.\  {\bf 828} (2011) 313
  [arXiv:1002.1722 [hep-th]].
	
\bibitem{D'Hoker:2009bc} 
  E.~D'Hoker and P.~Kraus,
  JHEP {\bf 1003}, 095 (2010)
  [arXiv:0911.4518 [hep-th]].
	
\bibitem{Domenech:2010nf}
  O.~Domenech, M.~Montull, A.~Pomarol, A.~Salvio and P.~J.~Silva,
  JHEP {\bf 1008} (2010) 033
  [arXiv:1005.1776 [hep-th]].
	
\bibitem{Dias:2013bwa}
  O.~J.~C.~Dias, G.~T.~Horowitz, N.~Iqbal and J.~E.~Santos,
  arXiv:1311.3673 [hep-th].
	
\bibitem{Gubser:2009qm}
  S.~S.~Gubser, C.~P.~Herzog, S.~S.~Pufu and T.~Tesileanu,
  Phys.\ Rev.\ Lett.\  {\bf 103} (2009) 141601
  [arXiv:0907.3510 [hep-th]].
	
\bibitem{Aprile:2011uq}
  F.~Aprile, D.~Roest and J.~G.~Russo,
  JHEP {\bf 1106} (2011) 040
  [arXiv:1104.4473 [hep-th]].
	
\bibitem{Gubser:2008px}
  S.~S.~Gubser,
  Phys.\ Rev.\ D {\bf 78} (2008) 065034
  [arXiv:0801.2977 [hep-th]].
	
\bibitem{Gregory:2009fj}
  R.~Gregory, S.~Kanno and J.~Soda,
  JHEP {\bf 0910} (2009) 010
  [arXiv:0907.3203 [hep-th]].
	
\bibitem{Ge:2011cw}
  X.~-H.~Ge,
  Prog.\ Theor.\ Phys.\  {\bf 128} (2012) 1211
  [arXiv:1105.4333 [hep-th]].
	
	
\bibitem{Ge:2012vp}
  X.~-H.~Ge, S.~F.~Tu and B.~Wang,
  JHEP {\bf 1209} (2012) 088
  [arXiv:1209.4272 [hep-th]].
	
	
\bibitem{Herzog:2008he}
  C.~P.~Herzog, P.~K.~Kovtun and D.~T.~Son,
  Phys.\ Rev.\ D {\bf 79} (2009) 066002
  [arXiv:0809.4870 [hep-th]].
	
	
	
\bibitem{Maeda:2008ir}
  K.~Maeda and T.~Okamura,
  Phys.\ Rev.\ D {\bf 78} (2008) 106006
  [arXiv:0809.3079 [hep-th]].
	
\bibitem{Yin:2013fwa}
  L.~Yin, D.~Hou and H.~-c.~Ren,
  arXiv:1311.3847 [hep-th].

\bibitem{Banerjee:2013maa}
  N.~Banerjee, S.~Dutta and D.~Roychowdhury,
  arXiv:1311.7640 [hep-th].
	

\bibitem{summers}
  LL.~T.~Summers, M.~W.~Guinan, J.~R.~Miller, and P.~A.~Hahn,
	IEEE Transactions on Magnetics, Vol. 27, No. 2 (1991)
	
	
\bibitem{Kim:2007fj}
  S.~Oh, D.~K.~Kim, C.~J.~Bae, H.~C.~Kim and K.~Kim,
  IEEE Transactions on Applied Superconductivity, Vol. 17, No. 2 (2007) 
	
\bibitem{Albash:2008eh}
  T.~Albash and C.~V.~Johnson,
  JHEP {\bf 0809} (2008) 121
  [arXiv:0804.3466 [hep-th]].
	
\bibitem{Kuang:2013oqa}
  X.~-M.~Kuang, E.~Papantonopoulos, G.~Siopsis and B.~Wang,
  Phys.\ Rev.\ D {\bf 88} (2013) 086008
  [arXiv:1303.2575 [hep-th]].
	
	
	
	
\bibitem{Albash:2009iq}
  T.~Albash and C.~V.~Johnson,
  Phys.\ Rev.\ D {\bf 80} (2009) 126009
  [arXiv:0906.1795 [hep-th]].
	
	
	
\bibitem{tinkham}
  M.~Tinkham,
  \textit{Introduction to Superconductivity}, 2nd edition, Dover; New York (1996).
	
	
	
	\end{thebibliography}
\end{document}